\documentclass	
  [superscriptaddress,amsmath,
    a4paper, 
    preprint, 
    floatfix
  ]
  {revtex4}
\usepackage[utf8x]{inputenc}
\usepackage[T1]{fontenc}
\usepackage{amsmath}
\usepackage{hyperref}
\usepackage{graphicx}
\usepackage{textcomp}
\graphicspath{{images/}}

\newcommand{\Ht}{\ensuremath{{^3\text{He}}}}
\newcommand{\Hf}{\ensuremath{{^4\text{He}}}}

\begin{document}


\title{Phase diagram of fluid phases in $\Ht$ -$\Hf$ mixtures}
\author{N. Farahmand Bafi}
\email{nimabafi@is.mpg.de}
\affiliation{Max-Planck-Institut f{\"u}r Intelligente Systeme, Heisenbergstr. 3, D-70569 Stuttgart, Germany}
\affiliation{IV. Institut f{\"u}r Theoretische Physik, Universit{\"a}t Stuttgart,Pfaffenwaldring 57, D-70569 Stuttgart, Germany}

\author{A. Macio\l ek}
\email{maciolek@is.mpg.de }
\affiliation{Max-Planck-Institut f{\"u}r Intelligente Systeme, Heisenbergstr. 3, D-70569 Stuttgart, Germany}
\affiliation{IV. Institut f{\"u}r Theoretische Physik, Universit{\"a}t Stuttgart,Pfaffenwaldring 57, D-70569 Stuttgart, Germany}
\affiliation{Institute of Physical Chemistry, Polish Academy of Sciences, Kasprzaka 44/52, PL-01-224 Warsaw, Poland}

\author{S. Dietrich}
\email{dietrich@is.mpg.de }
\affiliation{Max-Planck-Institut f{\"u}r Intelligente Systeme, Heisenbergstr. 3, D-70569 Stuttgart, Germany}
\affiliation{IV. Institut f{\"u}r Theoretische Physik, Universit{\"a}t Stuttgart,Pfaffenwaldring 57, D-70569 Stuttgart, Germany}

\begin{abstract}
Fluid parts of the phase diagram of $\Ht$ -$\Hf$ mixtures are obtained from a
mean-field
analysis of a suitable lattice gas model for binary liquid mixtures.
The proposed model takes into account the continuous rotational symmetry O(2) of the superfluid degrees of freedom associated with $\Hf$ and includes the occurrence of vacancies.
This latter degree of freedom allows the model to exhibit a vapor phase and hence can provide the theoretical framework to describe the experimental conditions
for
measurements of 
tricritical Casimir forces in $\Ht$ -$\Hf$ wetting films.
\end{abstract}


\maketitle


\section{INTRODUCTION}
 
\begin{figure}[t]
\includegraphics[width=85mm]{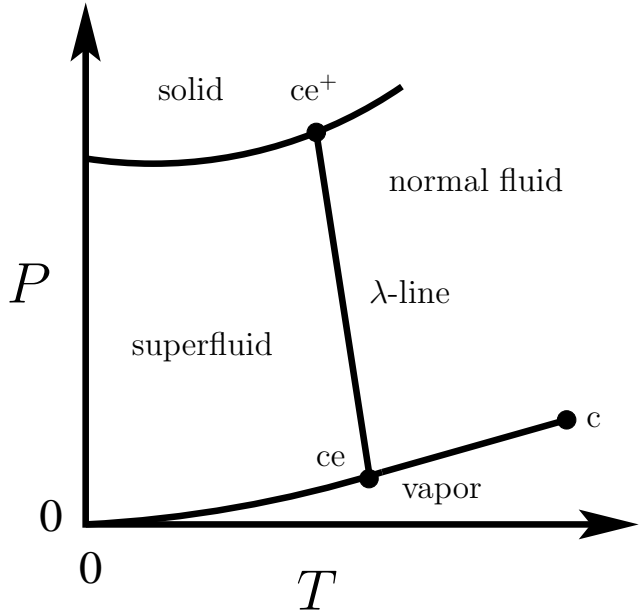}
\centering
\centering
\caption{
\linespread{1.0}
\small{Schematic
bulk phase
diagram of $\Hf$
exhibiting
the vapor, superfluid, normal fluid,
and solid phases.
The
liquid-vapor
critical point is denoted by c
whereas ce$^+$ and ce are critical
end points.
The
$\lambda$-line
is
the line of
second-order
phase transitions
between
the superfluid
and the normal fluid.
}}
\label{he4}
\end{figure}

\begin{figure}[t]
\includegraphics[width=85mm]{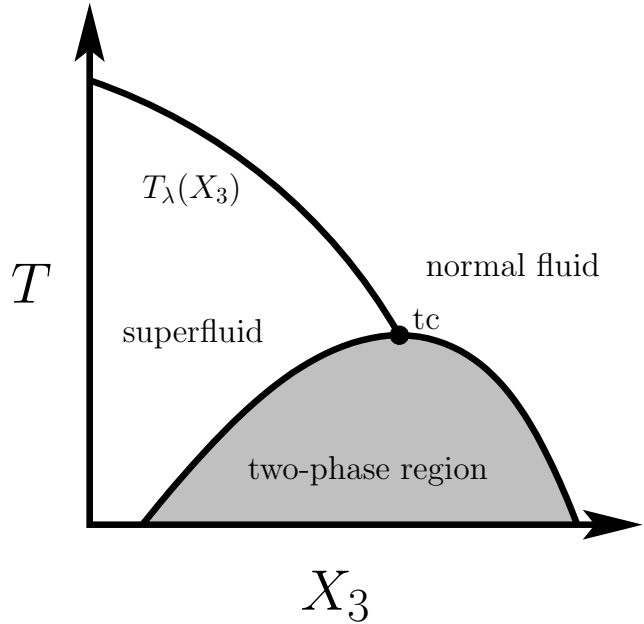}
\centering
\centering
\caption{
\linespread{1.0}
\small{Schematic
bulk phase
diagram of  $\Ht$ -$\Hf$ mixtures at fixed pressure. $X_3$ is the concentration of $\Ht$ and $T_{\lambda}(X_3)$ is the
line of continuous
superfluid
transitions, which
turn into first-order superfluid transitions at the tricritical point tc. Note that $T_{\lambda}(X_3)$ meets the two-phase region at its top. If $T_{\lambda}(X_3)$ would meet
the two-phase region below $T_{\text{tc}}$, this would imply that there is either a discontinuous phase transition between two normal fluid phases or between two superfluid phases,
which is not the case.
}}
\label{he43}
\end{figure}

\begin{figure}[t]
\includegraphics[width=85mm]{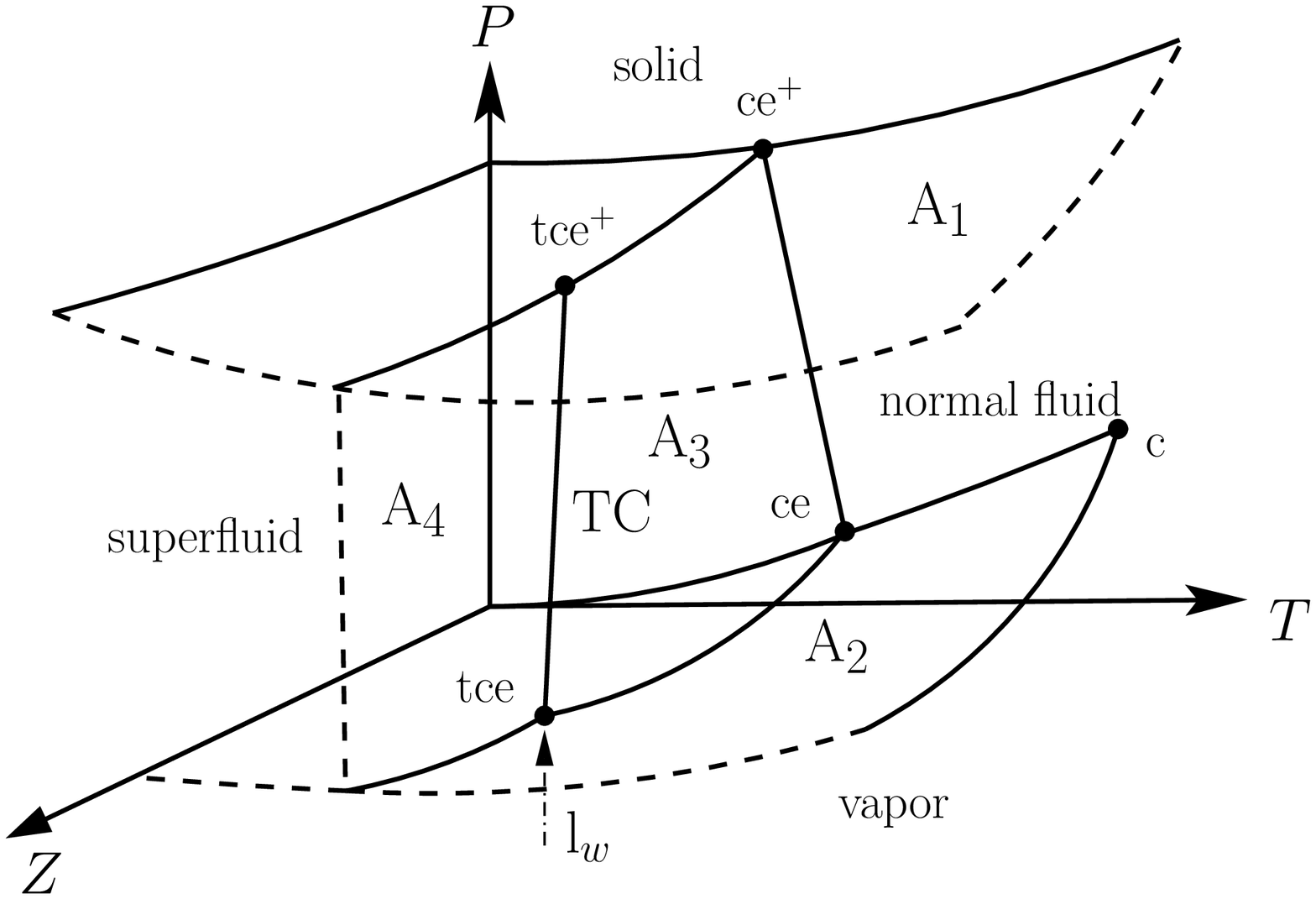}
\centering
\centering
\caption{
\linespread{1.0}
\small{Schematic phase diagram of $\Ht$ -$\Hf$ mixtures in the $(T,Z,P)$
space,
where
$Z=\exp(\mu_3/T)$
is the fugacity of $\Ht$
and $P$ is the pressure.
$A_1$ and $A_2$ are the
surfaces of the first-order
solid-liquid and vapor-liquid
phase transitions, respectively,
whereas $A_3$ and $A_4$ are the surfaces of
second- and first-order
phase transitions between
the normal fluid and
the
superfluid, respectively.
A$_3$ intersects A$_1$ and A$_2$ along a line of critical end points connecting ce$^+$ with tce$^+$
and ce with tce, respectively.
The surfaces $A_3$ and $A_4$ are separated by a
line of tricritical points TC
which
meets $A_1$ and $A_2$ at the tricritical end points
tce$^+$ and tce, respectively. $A_2$ terminates at
a
line of critical points, starting from c
in
the plane
$Z=0$.
The phase diagram in
the plane
$Z=0$ is the same as the one in Fig.~\ref{he4}.
The dashed lines have no physical meaning; they indicate that the corresponding surface continues. The arrow
l$_w$ indicates the thermodynamic path
along which tricritical
end point
wetting
occurs.
}}
\label{3dschematic-fig}
\end{figure}

Binary mixtures of the helium isotopes $\Ht$ and $\Hf$ exhibit a very rich phase behavior due to
the presence of pronounced
quantum effects.
For example, below a certain threshold value of the pressure the zero-point fluctuations of the helium atoms demolish the solid phase. Accordingly, the liquid phase persists down to temperature $T=0$.
The solid phase
forms
only at high pressures, whereas for sufficiently low pressures
and $T>0$
helium
forms the
vapor phase.
The bulk phase diagram of $\Hf$ is shown schematically in Fig.~\ref{he4}.
The liquid phase can be either
a
normal fluid or superfluid.
These two fluid phases
are separated by a line of 
second-order
phase transitions,
which is called $\lambda$-line.
This line
terminates at the critical end points ce$^+$ and ce at
the liquid-solid and liquid-vapor coexistence lines,
respectively.
The liquid-vapor coexistence line terminates at the critical point c.
\\
Adding $\Ht$ atoms to
the pure $\Hf$ liquid, dilutes the $\Hf$
carriers of superfluidity and thus lowers the
critical
temperature of the superfluid transition.
(Superfluid
transitions of $\Ht$ atoms
occur
at very low temperatures,
which
are not considered here.)
Beyond a certain dilution due to $\Ht$ atoms the superfluid transition turns into a first-order phase transition; this occurs at a tricritical point tc.
The schematic phase diagram of
$\Ht$ -$\Hf$
mixtures at fixed pressure is shown in Fig.~\ref{he43}.
The
transition
temperature
$T_{\lambda}$
of the
second-order
phase
transition
to
the
superfluid phase depends on the concentration
$X_3$
of $\Ht$ atoms.
For temperatures
below
the tricritical point tc,
the mixture
undergoes
a
first-order superfluid-normal phase transition which is accompanied by a two-phase region.
\\
The schematic phase diagram of $\Ht$ -$\Hf$ mixtures in the $(T,Z,P)$ space, where $Z$ is the fugacity of $\Ht$, is shown in Fig.~\ref{3dschematic-fig}~\cite{1922}.
In the plane $Z=0$, i.e.,
in the case of pure $\Hf$,
the phase diagram is the same as the one in Fig.~\ref{he4}.
$A_1$ and $A_2$ are the
surfaces of first-order
solid-liquid and vapor-liquid
transitions, respectively,
whereas $A_3$ and $A_4$ are the surfaces of
second- and first-order
phase transitions,
respectively,
between the superfluid and the normal fluid.
Accordingly, $A_3$ and $A_4$ are separated by a line TC of tricritical points, which terminates
at
the tricritical end points tce$^+$ and tce. The points tce$^+$ and ce$^+$ as well as
tce and ce are connected by lines of critical end points on $A_1$ and $A_2$, respectively. The surface $A_4$ intersects the surfaces $A_1$ and $A_2$ along triple lines
of three-phase coexistence between the solid and the two liquid phases and the vapor and the two liquid phases, respectively.

\textit{Classical}
lattice models
have turned out to
successfully describe the essential features of the phase diagram of binary
liquid
mixtures.
Such a model for describing the phase diagram of $\Ht$ -$\Hf$ mixtures near the tricritical point was first introduced and studied by Blume, Emery, and Griffiths (called the BEG model)~\cite{beg}.
In this
classical
spin-1 model, the superfluid order parameter
is mimicked
by two discrete values;
the remaining possible
value for the state variable indicates
whether a lattice site is occupied by a $\Ht$ atom
instead of a $\Hf$ atom.
Since this interpretation of the spin-1 model does not allow for vacancies, it does not exhibit a vapor phase. Furthermore, due to the discrete values assigned to the superfluid order parameter, this model does not capture the actual complex
character of the superfluid order parameter.
Another interpretation of the BEG model is to allow for vacant sites in a
\textit{classical}
binary
liquid
mixture of species A and B, which leads to the formation of
an A-rich liquid, a B-rich liquid, a mixed fluid phase, and a vapor phase.
Such a model
has been
used to study the condensation and
the
phase separation in binary liquid mixtures~\cite{lajzerowicz,dietrich-latz,dietrich-getta}.
The reduced phase diagrams of ternary mixtures have also been studied within this model~\cite{mukamel}.

Further improvements in the theoretical description of the phase diagrams of $\Ht$ -$\Hf$ mixtures have been
achieved
by
enriching the classical spin-$1/2$ model (i.e.,
without
vacancies) by
a continuous value for the superfluid order
parameter. 
Although this model takes into account the continuous O(2) symmetry of the superfluid order parameter, it does not incorporate the occurrence of a vapor phase.
Such a model with no vacancies and O(2) symmetry of the superfluid order parameter, is given by the so-called, vectorized BEG (VBEG) model which has been
proposed and studied in two dimensions (d=2) by Cardy and Scalapino~\cite{2dvbeg1} and, independently, by Berker and Nelson~\cite{2dvbeg2}.
More recently it has been investigated in d=3 within mean-field theory and by Monte Carlo simulations~\cite{vbeg3}.


In order to be able to study wetting films
in $\Ht$ -$\Hf$ mixtures
which have been used to analyze
experimentally the tricritical
Casimir effect~\cite{indirect},
the theoretical description of $\Ht$ -$\Hf$ mixtures requires to take into account the occurrence of a vapor phase.
Tricritical
Casimir forces acting on the
liquid-vapor interface
of $\Ht$ -$\Hf$
wetting
films arise due to the confinement of the 
tricritical
fluctuations of the superfluid order parameter and of the
composition
near the tricritical point of the mixture. The considerable interest in this subject has been triggered both by theoretical predictions~\cite{1922,1886}
and by experiments
in which superfluid wetting films ($\Hf$~\cite{chan4,ganshin} and $\Ht$ -$\Hf$~\cite{indirect}) were used to provide first reliable evidences for critical Casimir forces.
Specifically,
concerning tricriticality
a $\Ht$ -$\Hf$ mixture was prepared in
a
thermodynamic state of
the
vapor phase, close to coexistence with the liquid phase. Upon decreasing
undersaturation (see
the thermodynamic path l$_w$ in Fig.~\ref{3dschematic-fig}), a complete wetting
film was grown at the plates of capacitors, the equilibrium thickness of which could be determined very accurately from capacitance measurements. From the balance of the
effective
forces acting on the depinning liquid-vapor interface such as to thicken or to thinnen the film, the universal scaling function of the
tricritical
Casimir force was determined.

The, at present,
only available corresponding theoretical analysis~\cite{maciolek-gambassi-dietrich} of
the
behavior of the tricritical Casimir scaling functions
describing the $\Ht$ -$\Hf$ wetting film thicknesses, employs the VBEG model without vacancies, and thus does not incorporate the vapor phase.
Within this simplified approach, the wetting
films have
been modeled by a slab geometry with
the
boundaries introduced by fiat
and not
of via
the actual
self-consistent formation as a wetting film. Therefore, it is an open question how the critical Casimir
forces emerge in the $\Ht$ -$\Hf$ wetting films when
the system is brought
towards
the critical or the tricritical
end point, i.e., approaching
liquid-vapor coexistence
from the vapor side.
The present
bulk
analysis is a prerequisite of such investigations.

The model proposed here is a
classical
spin-1 model, including the continuous O(2) symmetry of the superfluid order parameter, which does allow for vacant sites and therefore exhibits a vapor phase if the number of vacant sites is
sufficiently large.
The phase diagrams of this model are obtained within
mean-field
theory. Since there are three order parameters
(i.e., the number
densities of $\Ht$ and $\Hf$ as well as the order parameter corresponding to the superfluid transition), the phase diagrams of the proposed model exhibit a rich diversity of topologies.
The main difficulty of the present study resides in extracting from a high dimensional parameter space the range of parameters for which the
phase diagrams have the topology corresponding to
the one of
the actual $\Ht$ -$\Hf$ mixture.
In the next section we introduce the model and continue by obtaining various features of the phase diagram. We close with a summary and conclusions. 


\section{THE MODEL}
We consider a three--dimensional
(d\,=\,3)
simple cubic lattice with lattice spacing $a=1$. The lattice sites $\{i\,|\,i=1,...,\mathcal{N}\}$ are occupied by either $\Ht$ or $\Hf$ or they
are unoccupied.
The Hamiltonian of this system is

\begin{equation}
\label{hamiltonian}
  \begin{split}
   \mathcal{H} & = -J_{44}N_{44}-J_{33}N_{33}-J_{34}N_{34}\\
               & \quad -\mu_{4}N_{4} -\mu_{3}N_{3} -J_{s}\tilde{N}_{44}-\mathbf{H}\cdot\tilde{\mathbf{N}}_{4}\text{,}
   \end{split}
\end{equation}
where $N_{\text{m} \text{n}}$, with $\text{m},\text{n}\in \{3, 4\}$, denotes the number of pairs of nearest neighbors of species m and n on the lattice
sites, $N_{\text{m}}$
denotes the number of
atoms of
species m,
and $-J_{s}\tilde{N}_{44}$
denotes the sum of the interaction energy between the superfluid degrees of freedom
$\Theta_{i}$ and $\Theta_{j}$
associated with the nearest--neighbor pairs $\langle i,j\rangle$ of $\Hf$ with
$J_s$ as the
corresponding
interaction strength. $ J_{33}$, $ J_{44}$, and $ J_{34}$ describe the effective interactions between the three types of pairs of He isotopes.
The
$\Ht$- $\Ht$ and $\Hf$- $\Hf$
pair potentials between the isotopes are not quite the same due to the slight
differences
in
their
electronic states.
Moreover, the corresponding effective interactions differ due to the distinct statistics of the two isotopes.
The chemical potential of species m is denoted as $\mu_{\text{m}}$. $\mathbf{H}=(H_x,H_y)$ is the field
conjugate
to the superfluid degrees of freedom given
by the vector $(\cos\Theta_i,\sin\Theta_i)$, provided that the lattice site $i$ is occupied by a
$\Hf$ atom.

In order to proceed, we express $N_{\text{m}}$ and $N_{\text{m} \text{n}}$ in terms of occupation numbers of the lattice sites $\{i\}$. We associate with each lattice site $i$ an occupation variable $s_{i}$ which can take the
three values $ +1$, $-1$, or $0$,
where $+1$ means that
the lattice site is occupied by $\Hf$, $-1$ means the lattice site is occupied by $\Ht$, and $0$ means the lattice site is unoccupied. Accordingly one has

\begin{equation}
 \begin{split}
 &N_{4}=\frac{1}{2} \sum_{i} s_{i} (s_{i}+1)\equiv \sum_{i} p_i\text{,}\\
 &N_{3}=\frac{1}{2} \sum_{i} s_{i} (s_{i}-1)\text{,}\\
 &N_{44}=\frac{1}{4} \sum_{<i,j>} (s_{i} (s_{i}+1)s_{j} (s_{j}+1))\equiv \sum_{<i,j>} p_i p_j\text{,}\\
 &N_{33}=\frac{1}{4} \sum_{<i,j>} (s_{i} (s_{i}-1)s_{j} (s_{j}-1))\text{,}\\
 &N_{34}=\frac{1}{4} \sum_{<i,j>} (s_{i} (s_{i}+1)s_{j} (s_{j}-1)+ s_{i} (s_{i}-1)s_{j} (s_{j}+1) )\text{,}
 \end{split}
\end{equation}
where $\sum \limits_{<i,j>}$ denotes the sum over nearest neighbors. Using the above definitions one obtains

\begin{equation}
  \begin{split}
   \mathcal{H} & = -K\sum_{<i,j>}s_{i}s_{j}-J\sum_{<i,j>}q_{i}q_{j}-C\sum_{<i,j>}(s_{i}q_{j}+q_{i}s_{j})\\
     & \quad -\Delta_{-}\sum_{i}s_{i}-\Delta_{+}\sum_{i}q_{i}-J_{s}\sum_{<i,j>}p_{i}p_{j}\cos(\Theta_{i}-\Theta_{j})\\
     & \quad -H_x\sum_{i}p_{i}\cos\Theta_i-H_y\sum_{i}p_{i}\sin\Theta_i\text{,}
  \end{split}
\end{equation}
where 
\begin{equation}
\label{Nvector}
  \begin{split}
     & \sum_{<i,j>}p_{i}p_{j}\cos(\Theta_{i}-\Theta_{j})= \tilde{N}_{44}= \sum_{<i,j>}p_{i}p_{j}\begin{pmatrix}\cos\Theta_{i}\\ \sin\Theta_{i}\end{pmatrix}\cdot\begin{pmatrix}\cos\Theta_{j}\\ \sin\Theta_{j} \end{pmatrix},\\
     & \sum_{i}p_{i}(\cos\Theta_{i},\sin\Theta_{i})=\tilde{\mathbf{N}}_{4},
  \end{split}
\end{equation}
and
\begin{equation}
\label{coupling}
 \begin{split}
 &q_{i}=s_{i}^2\text{,}\\
 &p_{i}= \frac{1}{2} s_{i}(s_{i}+1)\text{,}\\
 &K=\frac{1}{4}(J_{44}+J_{33}-2J_{34})\text{,}\\
 &J=\frac{1}{4}(J_{44}+J_{33}+2J_{34})\text{,}\\
 &C=\frac{1}{4}(J_{44}-J_{33})\text{,}\\
 &\Delta_{-}=\frac{1}{2}(\mu_{4}-\mu_{3})\text{,}\\
 &\Delta_{+}=\frac{1}{2}(\mu_{4}+\mu_{3})\text{,}
 \end{split}
\end{equation}
and $\Theta_{i}\in [0,2\pi]$ represents the superfluid degree of freedom at the lattice site i, provided it is occupied by $\Hf$.


\section{MEAN-FIELD THEORY}

In this section we apply mean-field
theory
to the above model. This approximation follows from a variational method based upon approximating the total equilibrium density
matrix by a product of density matrices associated with each lattice site~\cite{chaikin}.

Due to the variation principle, 
the free energy $F$ obeys the following inequality:
\begin{equation}
\label{var-principle}
F \leq \phi=\hat{T}r( \rho \mathcal{H}) + (1/\beta) \hat{T}r(\rho \ln \rho)\text{,}
\end{equation}
where $\rho$ is any trial density matrix
with $\hat{T}r(\rho)=1$,
with respect to which $\phi$ on the rhs of
Eq.~(\ref{var-principle})
should be minimized in order to obtain the best approximation;
\begin{equation}
\label{trace-formula}
\hat{T}r=\sum_{s_1=\pm 1,0}\int_0^{2\pi}\text{d}\Theta_1 \cdot... \cdot\sum_{s_\mathcal{N}=\pm 1,0}\int_0^{2\pi}\text{d}\Theta_\mathcal{N}
\end{equation}
denotes the trace
and $\beta=1/T$ where $T$ is the temperature times $k_\text{B}$.
The mean-field approximation
assumes that any lattice site experiences the same mean field generated by its neighborhood so that the total density matrix will be the product of
the density matrices corresponding to each lattice site: 
\begin{equation}
\label{RHO}
 \rho=\prod_{i} \rho _{i},
\end{equation}
with

\begin{equation}
Tr\rho_i=\sum_{s_i=\pm 1,0}\int_0^{2\pi}\text{d}\Theta_i \rho_i(s_i,\Theta_i)=1.
\end{equation}
For homogeneous bulk systems the local density matrix is independent of the site.

The variational mean-field free energy per site for the Hamiltonian introduced in the previous section is (with $\cos(\Theta_i-\Theta_j)=\cos\Theta_{i}\cos\Theta_{j}+\sin\Theta_{i}\sin\Theta_{j} $)
\begin{equation}
\label{phi1}
 \begin{split}
  \frac{\phi}{\mathcal{N}} & = - \frac{z}{2}[K\langle s_{i}\rangle^2+J\langle q_{i}\rangle^2+2C\langle q_{i}\rangle\langle s_{i}\rangle\\
                 &   \quad +J_s(\langle p_{i}\cos\Theta_{i}\rangle^{2}+\langle p_{i}\sin\Theta_{i}\rangle^{2})]\\
                 &   \quad  -\Delta_{-}\langle s_{i}\rangle-\Delta_{+}\langle q_{i}\rangle -H_x\langle p_{i}\cos\Theta_{i}\rangle-H_y\langle p_{i}\sin\Theta_{i}\rangle\\
                 &    \quad + (1/\beta) Tr(\rho_{i} \ln\rho_{i})\text{,}
  \end{split}
\end{equation}
where $\mathcal{N}$ is the total number of sites and $z$ is the coordination number of the lattice ($z=2d$, where $d$ is the spatial dimension of the system; here
$z=6$), and
$\langle...\rangle=Tr(\rho_i...)$
denotes the thermal average, taken with the trial density matrix $\rho_{i}$ associated with the lattice site i.

Minimizing
the variational function $\phi/\mathcal{N}$
with respect to $\rho_{i}$ renders
the best
normalized
functional form
of $\rho_{i}$.
There are two approaches to find the variational minima.
In the first approach one parametrizes the density matrix $\rho_i$ in terms of the order parameters of the phase transitions
and minimizes $\phi/\mathcal{N}$ with respect to the coefficients multiplying these order parameters. In the second approach one treats $\rho$ itself as a variational
function and minimizes $\phi/\mathcal{N}$ with respect to
it \cite{chaikin}.
We follow the second approach and calculate the functional derivative of $\phi/\mathcal{N}$
in Eq.~(\ref{phi1})
with respect to $\rho_{i}(s_i,\Theta_i)$
using $\frac{\delta \rho_{i}(s_i,\Theta_i)}{\delta \rho_{j}(s_j,\Theta_j)}=\delta (\Theta_i-\Theta_j)\delta_{s_i,s_j}$,
and equate it to the Lagrange multiplier $\eta$
corresponding to the
constraint
$Tr(\rho_i)=1$
\begin{equation}
\begin{split}
\label{EqEta}
\eta &= \frac{\delta (\phi/\mathcal{N})}{\delta \rho_{i}(s_i,\Theta_i)}\\
     &= - z [K\langle s_{i}\rangle s_{i}+J\langle q_{i}\rangle q_{i}+C( q_{i}\langle s_{i}\rangle+\langle q_{i}\rangle s_{i})\\
                 &   \quad +J_s(\langle p_{i}\cos\Theta_{i}\rangle p_{i}\cos\Theta_{i}+\langle p_{i}\sin\Theta_{i}\rangle p_{i}\sin\Theta_{i})]\\
                 &   \quad  -\Delta_{-} s_{i}-\Delta_{+} q_{i} -H_x p_{i}\cos\Theta_{i}-H_y p_{i}\sin\Theta_{i}\\
                 &    \quad + (1+\ln \rho_i)/\beta\text{.}
\end{split}
\end{equation}
Equation~(\ref{EqEta}) can be solved for $\rho_{i}(s_i,\Theta_i)$:
\begin{equation}
\rho_{i}=e^{\beta \eta -1-\beta h_{i}}\text{,}
\end{equation}
where
\begin{equation}
\label{eq h}
 \begin{split}
   h_{i}(s_i,\Theta_i)=& -s_{i}(kX+cD+\Delta_{-})-q_{i}(jD+cX+\Delta_{+})\\
	 & -p_{i}((j_{s}M_{x}+H_x)\cos\Theta_{i}+(j_{s}M_{y}+H_y)\sin\Theta_{i})
 \end{split}
\end{equation}
is the single-site Hamiltonian
in which the coupling constants are rescaled as $j=zJ$, $c=zC$, $k=zK$, $j_{s}=zJ_{s}$ and
where
the following order parameters
are introduced:
\begin{equation}
\label{definition of order parameters}
 \begin{split}
   & X:=\langle s_{i}\rangle\text{,}\\
   & D:=\langle q_{i}\rangle\text{,}\\
   & M_{x}:=\langle p_{i}\cos\Theta_{i}\rangle\text{,}\\
   & M_{y}:=\langle p_{i}\sin\Theta_{i}\rangle\text{,}  
 \end{split}
\end{equation}
which in the
bulk
are independent of $i$. In accordance with Eq.~(\ref{Nvector}) one has
$\langle\tilde{\mathbf{N}}_{4}\rangle=\mathcal{N}\mathbf{M}$.
The normalization $Tr(\rho_i)=1$ yields
\begin{equation}
e^{-\beta \eta +1}=Tr(e^{-\beta h_{i}})
\end{equation}
so that
\begin{equation}
\label{rho}
\rho_{i}=\frac{e^{-\beta h_{i}}}{Tr(e^{-\beta h_{i}})}\text{,}
\end{equation}
where $h_{i}$ is given by Eq.~(\ref{eq h}).

The order parameters defined in Eq.~(\ref{definition of order parameters})
allow one to determine the number densities $X_4=\frac{\langle N_4\rangle}{\mathcal{N}}=\frac{D+X}{2}$ and $X_3=\frac{\langle N_3\rangle}{\mathcal{N}}=\frac{D-X}{2}$
so that $X=(\langle N_4 \rangle-\langle N_3 \rangle)/\mathcal{N}=X_4-X_3$ is the difference of the number densities and $D=(\langle N_4 \rangle+\langle N_3 \rangle)/\mathcal{N}$
is the total number density. The concentration of $\Hf$ and $\Ht$ is $\frac{\langle N_4 \rangle}{\langle N_4 \rangle+\langle N_3 \rangle}\equiv\mathcal{C}_4=\frac{D+X}{2D}=X_4/D$
and $\frac{\langle N_3 \rangle}{\langle N_4 \rangle+\langle N_3 \rangle}\equiv\mathcal{C}_3=\frac{D-X}{2D}=X_3/D$, respectively.
$M_{x}$ and $M_{y}$ are the components of the
two-dimensional superfluid order parameter $\mathbf{M}=(M_{x}, M_{y})$ with $M:=\sqrt{|\mathbf{M}|^2}=\sqrt{M_{x}^2+M_{y}^2}$.
The
equilibrium superfluid order parameter
$\mathbf{M}$ points
into the direction
of $\mathbf{H}$.
This follows from the principle of minimum free energy
together with the relation
$\frac{\partial F}{\partial\mathbf{H}}=-\mathbf{M}$, where $F$ is the free energy of the system, which implies that for fixed $T$, $\Delta_+$, and $\Delta_-$
one has $\text{d}F=-\text{d}\mathbf{H}\cdot\mathbf{M}$.
Thus for $\mathbf{H}$
with an orientation $\psi$, i.e.,
$\mathbf{H}=(H_x,H_y)= H (\cos\psi,\sin\psi)$ with $H:=\sqrt{|\mathbf{H}|^2}=\sqrt{H_{x}^2+H_{y}^2}$,
$\mathbf{M}$ points into the same direction,
i.e.,
$\mathbf{M}=(M_x,M_y)=M(\cos\psi,\sin\psi)$.
\\
Within
the aforementioned mean field
approximation the order parameters
$X(\Delta_{-},\Delta_{+},H,T)=Tr(\rho_i s_i)$, $D(\Delta_{-},\Delta_{+},H,T)=Tr(\rho_i q_i)$, and $M(\Delta_{-},\Delta_{+},H,T)$ (with the latter obtained from $M_x=Tr(\rho_i p_i \cos\Theta_i)$ and $M_y=Tr(\rho_i p_i \sin\Theta_i)$)
are given by
three coupled self-consistent equations:
\begin{equation}
\label{eqx}
 X=\frac{-W(X,D;\Delta_{-},\Delta_{+},H,T)+R(X,D;\Delta_{-},\Delta_{+},H,T)I_{0}(\beta j_{s}M+\beta H)}{1+W(X,D;\Delta_{-},\Delta_{+},H,T)+R(X,D;\Delta_{-},\Delta_{+},H,T)I_{0}(\beta j_{s}M+\beta H)}\text{,}
\end{equation}

\begin{equation}
\label{eqd}
 D=\frac{W(X,D;\Delta_{-},\Delta_{+},H,T)+R(X,D;\Delta_{-},\Delta_{+},H,T)I_{0}(\beta j_{s}M+\beta H)}{1+W(X,D;\Delta_{-},\Delta_{+},H,T)+R(X,D;\Delta_{-},\Delta_{+},H,T)I_{0}(\beta j_{s}M+\beta H)}\text{,}
\end{equation}
and
\begin{equation}
\label{eqm}
 M=\frac{
R(X,D;\Delta_{-},\Delta_{+},H,T)I_{1}(\beta j_{s}M+\beta H)}{1+W(X,D;\Delta_{-},\Delta_{+},H,T)+R(X,D;\Delta_{-},\Delta_{+},H,T)I_{0}(\beta j_{s}M+\beta H)}\text{,} 
\end{equation}
where $I_{0}$ and $I_{1}$ are modified Bessel functions
(see Sec.~9.6 in Ref.~\cite{abramowitz}).
The functions
$W(X,D;\Delta_{-},\Delta_{+},H,T)$ and  $R(X,D;\Delta_{-},\Delta_{+},H,T)$ are given by
\begin{equation}
\label{EqW}
W(X,D;\Delta_{-},\Delta_{+},H,T)=e^{\beta[(c-k)X+(j-c)D+\Delta_{+}-\Delta_{-}]}>0
\end{equation}
and
\begin{equation}
\label{EqR}
R(X,D;\Delta_{-},\Delta_{+},H,T)=e^{\beta[(c+k)X+(j+c)D+\Delta_{+}+\Delta_{-}]}>0
\end{equation}
so that $D>X$.
The equilibrium free energy $\phi(\Delta_{-},\Delta_{+},H,T)$ is given by 
\begin{equation}
 \label{free energy}
\begin{split}
\phi(\Delta_{-},\Delta_{+},H,T)= & \frac{k}{2}X^2+\frac{j}{2}D^2+cXD+\frac{j_{s}}{2}M^2\\
                               &+\frac{1}{\beta} \ln(1-D)\text{.}
\end{split}
\end{equation}
In the limit $\Delta_{+}\rightarrow +\infty$ both $W$ and $R$ diverge so that according to Eq.~(\ref{eqd}) one has $D(\Delta_{-},\Delta_{+}\rightarrow +\infty,H,T)\rightarrow 1$, i.e., all lattice
sites are occupied and the concentrations reduces to $\mathcal{C}_4=(1+X)/2$ and $\mathcal{C}_3=(1-X)/2$. With the explicit expressions in Eqs.~(\ref{EqW}) and (\ref{EqR}), in the limit $\Delta_{+}\rightarrow +\infty$,
Eqs.~(\ref{eqx}) and~(\ref{eqm}),
reduce to:
\begin{equation}
\label{limit1}
X=\frac{-e^{-\beta(2kX+2c+2\Delta_{-})}+I_{0}(\beta j_{s}M+\beta H)}{e^{-\beta(2kX+2c+2\Delta_{-})}+I_{0}(\beta j_{s}M+\beta H)},\qquad\Delta_+=\infty,
\end{equation}
and
\begin{equation}
\label{limit2}
M=\frac{I_{1}(\beta j_{s}M+\beta H)}{e^{-\beta(2kX+2c+2\Delta_{-})}+I_{0}(\beta j_{s}M+\beta H)},\qquad\Delta_+=\infty\text{.}
\end{equation}
Expressing $X$ in Eqs.~(\ref{limit1}) and (\ref{limit2}) in terms of $\mathcal{C}_4$ renders
\begin{equation}
\mathcal{C}_4=\frac{I_{0}(\beta j_{s}M+\beta H)}{e^{\beta(-\tilde{k}\mathcal{C}_4+\tilde{\Delta}_{-})}+I_{0}(\beta j_{s}M+\beta H)},\qquad\Delta_+=\infty\text{,}
\end{equation}
and 
\begin{equation}
M=\frac{I_{1}(\beta j_{s}M+\beta H)}{e^{\beta(-\tilde{k}\mathcal{C}_4+\tilde{\Delta}_{-})}+I_{0}(\beta j_{s}M+\beta H)},\qquad\Delta_+=\infty\text{,}
\end{equation}
where $\tilde{k}=4k$ and $\tilde{\Delta}_{-}=2(-\Delta_{-}+k-c)$.
For $H=0$ these equations have the same form as the corresponding ones in Ref.~\cite{vbeg3}, which
do not allow for vacant sites from outset.
Thus in the limit $\Delta_{+}\rightarrow +\infty$ and for $H=0$
our present more general results reduce to those of the more restricted model
studied before.

\section{PHASE DIAGRAM}

In this section we
determine
the phase diagram of the VBEG model within mean-field theory.
Although certain features of the phase diagram can be obtained analytically,
most parts of it can be determined only numerically.
In order to find the coexisting states of phase equilibria, one has to identify those distinct states $(X_\nu,D_\nu,M_\nu)$, which share
the same values for the chemical potentials and the pressure at
a common temperature.
The chemical potentials can be obtained by solving Eqs.~(\ref{eqx}) and (\ref{eqd})
together with Eqs.~(\ref{EqW}) and ~(\ref{EqR})
for $\Delta_{+}$ and $\Delta_{-}$:
\begin{equation}
\label{eqdelta+}
\begin{split}
 \Delta_{+}(X,D,M;H,T)\,= \,\,&\, \frac{T}{2}\ln(D^2-X^2)-T\ln(2(1-D))-cX-jD\\
	     & \,-\frac{T}{2}\ln(I_{0}(j_{s}M/T+H/T))
 \end{split}
\end{equation}
and
\begin{equation}
\label{eqdelta-}
 \Delta_{-}(X,D,M;H,T)\,=\,\,\frac{T}{2}\ln \frac{D+X}{D-X}-kX-cD-\frac{T}{2}\ln(I_{0}(j_{s}M/T+H/T))\text{.}
\end{equation}
Within the grand canonical ensemble the pressure is given by $\phi/\mathcal{N}=-P$. (Note that the sample volume is $V=\mathcal{N}a^3$, here with $a=1$).
According to Eqs.~(\ref{eqx}) - (\ref{eqm})
the order parameters of any state must fulfill the
relation
\begin{equation}
\label{equiM}
\frac{2M}{X+D}=\frac{M}{X_4}=\frac{I_{1}(\beta j_{s}M+\beta H)}{I_{0}(\beta j_{s}M+\beta H)}\text{,}
\end{equation}
which expresses $M$ in terms of $\frac{X+D}{2}=X_4$,
$T$, and $H$.
Depending on the value of the coupling constant $j_s$ the phase diagram exhibits
various
topologies.
\subsection{Phase diagram for a simple, normal liquid mixture: $j_s=0$}
\begin{figure}[t]
\includegraphics[width=95mm]{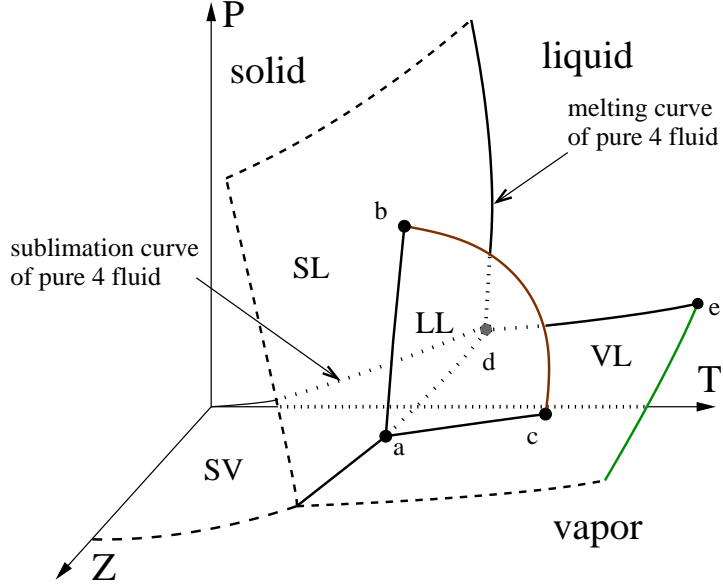}
\centering
\centering
\caption{
\linespread{1.0}
\small{
The schematic phase diagram for
$H=0$ and $j_s=0$ (i.\,e., without coupling between the superfluid degrees of freedom).
The phase diagram in the plane $Z=0$ is that of a one-component system
consisting of particles "4".
Upon
increasing the fugacity $Z$
of particles "3",
the transition lines in the plane $Z=0$ extend to form three distinct surfaces. The surface SL is
a
surface of first-order phase transitions between the solid and
the liquid phases.
The
transition surfaces between the vapor and the liquid phases, and
between
the solid and the liquid phases are denoted by VL and SV, respectively.
The surface VL terminates at
a
line of critical points (green line).
The critical point of the pure system of "4" particles is denoted by 'e'.
The liquid can be either
mixed or demixed.
Concerning the demixed
phases,
LL denotes the surface of first-order phase transitions between the phase rich in component 3
(large $Z$)
and the phase rich in the component 4
(small $Z$).
This surface terminates at
a
line of critical points (brown line), which meets the surfaces SL and VL at the critical end points 'b' and 'c', respectively.
The point
'a' is a quadruple point, whereas 'd'
is a triple point.
The lines a-b, a-c, and a-d are triple lines.
The
dashed
lines have no
physical meaning; they
indicate
that the
corresponding
surfaces continue.
}}
\label{3dA}
\end{figure}

For $j_s=0$ and $H=0$
there is no superfluid phase and $M$ is always zero (compare Eq.~(\ref{equiM}) with $I_1(y\rightarrow0)=\frac{1}{2}y$ and $I_0(y\rightarrow0)=1$). For $j_s=0$,
due to $I_0(0)=1$ the last term in Eq.~(\ref{eqdelta+}) and in Eq.~(\ref{eqdelta-}) drops out.
Thus the phase diagram will be that of a simple binary normal liquid mixture
of species 3 and 4,
similar to the ones
shown in Refs.~\cite{lajzerowicz,dietrich-latz,dietrich-getta}.
The
first-order demixing transitions occur at low temperatures, whereas at
high temperatures the liquid is mixed. The demixing transitions
terminate in a
line of critical points
which due to $\frac{{\partial}\phi}{{\partial}X}=-\Delta_{-}$ (see Eqs.~(\ref{phi1}) and (\ref{definition of order parameters})) are given by~\cite{chaikin}
\begin{equation}
\label{criticalCond}
\frac{\mathrm{d}\Delta_{-}}{\mathrm{d}X}|_{\Delta_{+},T}=\frac{\mathrm{d}^2\Delta_{-}}{\mathrm{d}X^2}|_{\Delta_{+},T}=0\text{,} \quad \frac{\mathrm{d}^3\Delta_{-}}{\mathrm{d}X^3}|_{\Delta_{+},T}>0\text{,}
\end{equation}
where $\frac{\mathrm{d^{\text{n}}}\Delta_{-}}{\mathrm{d}X^{\text{n}}}|_{\Delta_{+},T}$ denotes the $\text{n}^{\text{th}}$ total derivative of $\Delta_{-}$ 
(see Eq.~(\ref{eqdelta-})) with respect to $X$
at constant $\Delta_{+}$ and $T$.
Note that the independent variables are $(T,\Delta_{+},\Delta_{-})$.
Since $\Delta_{-}$ as given by Eq.~(\ref{eqdelta-}) depends on $D$, which for
$\Delta_{+}=const$
depends in turn implicitly on
$X$ via Eq.~(\ref{eqdelta+}), calculating the total
derivative
of $\Delta_{-}$ with respect to $X$
requires the knowledge of the partial
derivative
of $D$ with respect to $X$. Thus the first condition in Eq.~(\ref{criticalCond}) reads
\begin{equation}
\label{criticalCond1}
\frac{\mathrm{d}\Delta_{-}}{\mathrm{d}X}|_{\Delta_{+},T}=\frac{\mathrm{\partial}\Delta_{-}}{\mathrm{\partial}X}|_{\Delta_{+},T}+\frac{\mathrm{\partial}\Delta_{-}}{\mathrm{\partial}D}
\frac{\mathrm{\partial}D}{\mathrm{\partial}X}|_{\Delta_{+},T}=0\text{,}
\end{equation}
where $\frac{\partial\Delta_{-}}{\partial X}$ and $\frac{\partial\Delta_{-}}{\partial D}$ follow from Eq.~(\ref{eqdelta-}) and
where $\frac{\mathrm{\partial}D}{\mathrm{\partial}X}|_{\Delta_{+},T}$ is obtained by taking the derivative of Eq.~(\ref{eqdelta+}) with
respect to $X$ at fixed $\Delta_{+}$ and $T$ and by solving for $\frac{\mathrm{\partial}D}{\mathrm{\partial}X}|_{\Delta_{+},T}$.
Accordingly, the
first condition in Eq.~(\ref{criticalCond}) leads to a quadratic equation:
\begin{equation}
\label{T-quad}
T^2+a_1T+a_0=0
\end{equation}
with
\begin{equation}
\begin{split}
 &a_1=-D (-2 c X+j+k)+X (k X-2 c)+D^2 j\text{,}\\
 &a_0=(D-1) (D^2-X^2) \left(c^2-j k\right).
\end{split}
\end{equation}
Equation~(\ref{T-quad}) renders as solution two branches $T_{1,2}(X,D)$.
Similarly, the second condition in Eq.~(\ref{criticalCond}) leads to an equation $G(X,D,T)=0$ where,
due to the first condition, $T=T_{1,2}(X,D)$.
Thus it takes the form
$G(X,D,T_{1,2}(X,D))=:g_{1,2}(X,D)=0$.
Therefore,
for a given value $D^{(0)}$ of $D$,
the solution of $g_{1,2}(X,D^{(0)})=0$ (which must be solved numerically)
renders $X(D_{1,2}^{(0)})=X_{1,2}^{(0)}$ so that at $T_{1,2}^{(0)}=T_{1,2}(X_{1,2}^{(0)},D_{1,2}^{(0)})$ the model exhibits
a critical point,
provided
the condition $\frac{\mathrm{d}^3\Delta_{-}}{\mathrm{d}X^3}|_{\Delta_{+},T}>0$
is fulfilled.
This latter condition and the physical
constraints
$T>0$, $P>0$, and $D>|X|$ exclude one of the two branches of $T_{1,2}(X,D)$.
Thus for
various values of $D$ one obtains a set of points $\{(D,X(D),T(X(D),D))\}$, which
forms
a line of critical points in the space spanned by $(X,D,T)$.
According to Eqs.~(\ref{eqdelta+}) and~(\ref{eqdelta-}), the set $\{(D,X,T)\}$ can be transformed to
the
set $\{(\Delta_{+}(D,X;T),\Delta_{-}(D,X;T),T)\}$, which
yields a
line of critical points in the space spanned by $(\Delta_{+},\Delta_{-},T)$.
This line ends at the liquid-vapor coexistence surface forming a critical
end point
(see Fig.~\ref{3dA}).

 \begin{figure}[t]
\includegraphics[width=85mm]{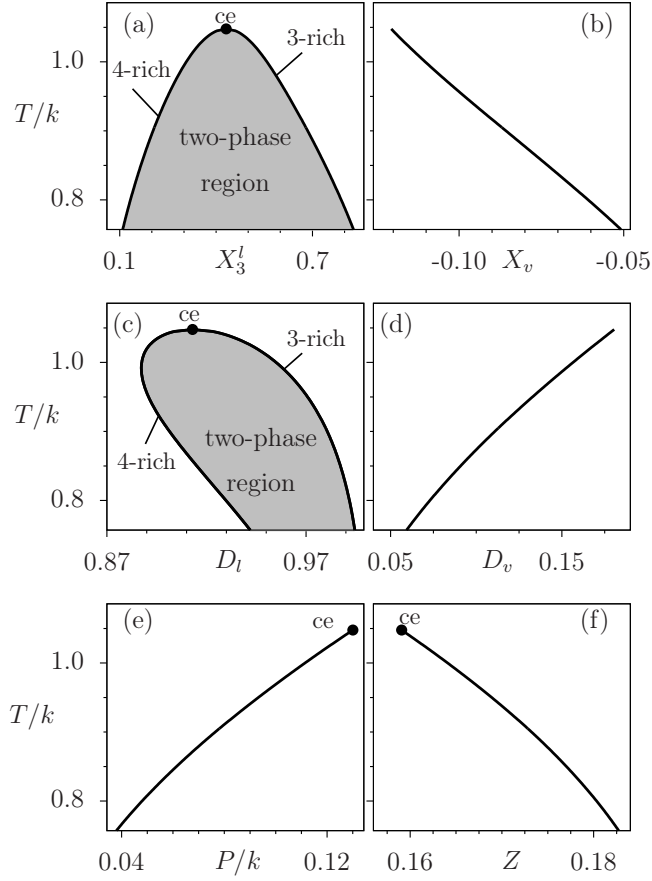}
\centering
\centering
\caption{
\linespread{1.0}
\small{
Phase diagrams for the coupling constants $(c/k, j/k, j_s/k)=(1, 5.714, 0)$ and $H=0$.
Along the triple line of three-phase coexistence
(see line a-c in Fig.~\ref{3dA})
the figures show the first-order
demixing transitions
of
the liquid at coexistence with the vapor phase
(a) in the $(X^{l}_3$, $T)$ plane, with
$X^{l}_3=\langle N_3\rangle_l/\mathcal{N}$
corresponding to the 3-particles in the liquid phase, (b) in the $(X_{v}$, $T)$ plane at
coexistence with the two liquid phases, where $X_v=(\langle N_4\rangle_v-\langle N_3\rangle_v)/\mathcal{N}$ in the vapor phase, (c) in the $(D_{l}$, $T)$ plane,
where $D_l=(\langle N_4\rangle_l+\langle N_3\rangle_l)/\mathcal{N}$ in the liquid phase, and (d) in the $(D_{v}$, $T)$ plane,
where
$D_v=(\langle N_4\rangle_v+\langle N_3\rangle_v)/\mathcal{N}$
in the vapor phase. The indices
'$l$' and '$v$' refer to the values of the order parameters in the liquid and
in the vapor phase,
respectively.
The critical end point ce here corresponds to the point c in Fig.~\ref{3dA}.
Panels (e) and (f) show
the dependences of the pressure $P$ and of the fugacity $Z=\exp(\mu_3/T)$ on the temperature along the triple line a-c in
Fig.~\ref{3dA}.
The two liquid states become identical at the critical
end point
ce at $T_{\text{ce}}/k=1.947$, above which the liquid is
mixed.
The coexisting liquid and vapor phases at ce are 
$(X^{\text{ce}}_l, D^{\text{ce}}_l, M^{\text{ce}}_l)=(0.050, 0.913, 0)$ and $(X^{\text{ce}}_v, D^{\text{ce}}_v, M^{\text{ce}}_v)=(-0.120, 0.180, 0)$,
respectively. The transitions
between
the vapor phase
and the liquid
phases
are always
first order.
According to (e) and (f), along a-c both $P$ and $Z$ vary as function of $T$, with the requirement of staying in coexistence with the vapor phase.
This implies that the white domains in (a) and (c) are not projections of a three-dimensional surface, given by the equation of state, onto the $(T, X_3)$ and
$(T, D)$ plane, respectively. The black lines provide only the $T$-dependence of $X_3$ and $D_l$ along the line a-c, which contains two branches. Similar remarks hold for Figs.~\ref{t2}-\ref{t5}.
}}
\label{t1}
\end{figure}

The schematic phase diagram for $j_s=0$ in the $(T,Z,P)$ space is shown in Fig.~\ref{3dA}, with $Z=\exp(\mu_3/T)$.
There
are four
surfaces separating
various
phases: the surface
SL
of first-order phase transitions between the solid and
the liquid phases, the surface
VL of first-order phase transitions
between the vapor and the liquid phases, the surface
SV
of first-order phase transitions between the solid and the liquid phases,
and the surface
LL
of first-order phase transitions between the phase rich in component 3 and the phase rich in the component 4.
This latter surface terminates at
a
line of critical points (brown line), and
VL terminates at
a
line of critical points (green line).

In Fig.~\ref{t1},
the demixing transitions at coexistence with the vapor phase
(see the line connecting the points 'a' and 'c' in Fig.~\ref{3dA})
are shown
for the coupling constants
chosen as
$(c/k, j/k, j_s/k)=(1, 5.714, 0)$.
Along
this triple
line of
first-order
liquid-liquid transitions at coexistence with the vapor phase, three
thermodynamic
states with
distinct number
densities and concentrations coexist.
The
values of the order parameters of these three states are shown in Figs.~\ref{t1}(a)-(d).
The
corresponding values of the pressure
$P/k$
and of the fugacity, $Z=\exp(\mu_3/T)=\exp(\frac{\mu_3}{k}\frac{1}{T/k})$ of the component 3 of the mixture ($\mu_3=\Delta_{+}-\Delta_{-}$ and $T$ are rescaled by the coupling constant $k$) are shown
in Figs.~\ref{t1}(e) and (f),
respectively.
The vapor phase is characterized by a small value $D_v$ of the order parameter $D$, whereas
a
large value $D_l$ of the density order parameter $D$ corresponds to the liquid state.
In Fig.~\ref{t1}, at fixed temperatures below the critical end point
(ce)
(which is denoted as 'c' in Fig.~\ref{3dA}),
three values for $X_3$,
i.e., two values $X_3^l$ for $X_3$ in the liquid
phases
(Fig.~\ref{t1}(a)) and one value for the vapor phase ($X_v$ in Fig.~\ref{t1}(b)),
and three values for $D$ (Figs.~\ref{t1} (c)-(d)) characterize the three states which share the same values of the pressure (Fig.~\ref{t1}(e)) and of the chemical potentials (and thus the fugacity,
Fig.~\ref{t1}(f)).
At
$T_{\text{ce}}/k=1.947$
the two liquid states
merge into
a single state
with $(X^{\text{ce}}_l,D^{\text{ce}}_l,M^{\text{ce}}_l)=(0.050, 0.913, 0)$, which coexists with the vapor state
characterized by $(X^{\text{ce}}_v,D^{\text{ce}}_v,M^{\text{ce}}_v)=(-0.120, 0.180, 0)$.
For $T>T_{\text{ce}}$ the liquid is mixed. The transitions between the vapor and
the
liquid phases are always
first order,
above and below $T_{\text{ce}}$.

\subsection{Phase diagram including the superfluid phase: $j_s>0$}

For $j_s>0$ the model exhibits superfluid transitions, which can be either
first or second
order.
In order to find the surface of second-order
phase transitions to the superfluid phase (see A$_3$ in Fig.~\ref{3dschematic-fig}), we
introduce
the appropriate thermodynamic
potential $A$ as the Legendre transform of $\phi$:
\begin{equation}
 A(\Delta_{-},\Delta_{+},M,T)=\phi(\Delta_{-},\Delta_{+},H(\Delta_{-},\Delta_{+},M,T),T)-MH(\Delta_{-},\Delta_{+},M,T)\text{,}
\end{equation}
where, according to Eq.~(\ref{hamiltonian}), $\frac{\partial\phi(\Delta_{-},\Delta_{+},M,T)}{\partial H}=M$
which implicitly renders $H=H(\Delta_{-},\Delta_{+},M,T)$ so that $\frac{\partial A(\Delta_{-},\Delta_{+},M,T)}{\partial M}=-H(\Delta_{-},\Delta_{+},M,T)$.
In order to determine
$H(\Delta_{-},\Delta_{+},M,T)$
we use
Eq.~(\ref{equiM}).
Because we are interested in the phase diagram
for
$H=0$, we replace the right hand side of
Eq.~(\ref{equiM})
by its approximation linear in $H$:
\begin{equation}
\label{expandM}
\frac{2M}{X+D}=\frac{I_1\left(j_{s}M/T\right)}{I_0\left(j_{s}M/T\right)}+
\frac{H}{T}\frac{I_1^{'}\left(j_{s}M/T\right)I_0\left(j_{s}M/T\right)-I_0^{'}\left(j_{s}M/T\right)I_1\left(j_{s}M/T\right)}{I_0^2\left(j_{s}M/T\right)}.
\end{equation}
Solving this equation for $H$ (using $I_0^{'}=I_1$, $I_1^{'}=(I_0+I_2)/2$, and $I_2(a)=I_0(a)-\frac{2}{a}I_1(a)$) leads to
 
\begin{widetext}
\begin{equation}
\label{equH}
\begin{split}
H&=\frac{I_1(j_{s}M/T)}{M/T}\frac{2M I_0(j_{s}M/T)-(X+D)I_1(j_{s}M/T)}{I_0^2(j_{s}M/T)+I_0(j_{s}M/T)I_2(j_{s}M/T)-2I_1^2(j_{s}M/T)}\\
 &=\frac{-j_sT I_{1}(j_{s}M/T)[-2M I_{0}(j_{s}M/T) + (D+X) I_{1}(j_{s}M/T)]}{2[-j_sM (I_{1}(j_{s}M/T))^2+ I_{0}(j_{s}M/T) [T I_{1}(j_{s}M/T) +j_sM I_{2}(j_{s}M/T)]]}.
  \end{split}
\end{equation}
\end{widetext}
Due to $\frac{\partial A}{\partial M}=-H$ the conditions for the critical points, where $M$ vanishes continuously (see $\text{A}_3$ in Fig.~\ref{3dschematic-fig}), are
(compare Eq.~(\ref{criticalCond}))
\begin{equation}
\label{conditioncritical}
 \frac{dH}{dM}|_{\Delta_{+},\Delta_{-},T}=\frac{d^2 H}{dM^2}|_{\Delta_{+},\Delta_{-},T}=0 \text{, } \quad \frac{d^3 H}{dM^3}|_{\Delta_{+},\Delta_{-},T}>0\text{,}
\end{equation}
with all total
derivatives to be taken at $M=0$ and
at constant $\Delta_{+}$, $\Delta_{-}$, and $T$
(compare Eq.~(\ref{criticalCond1})).
Note that the independent variables are $(T,\Delta_{+},\Delta_{-})$.
According to Eq.~(\ref{equH}),
calculating the total derivatives of $H$ with respect to $M$ requires the expression
for
$\frac{\partial H}{\partial M}$ and the knowledge of the partial derivatives of $X$ and $D$ with respect to $M$.
These
latter
ones are obtained by taking
the
partial derivatives of Eqs.~(\ref{eqdelta+}) and (\ref{eqdelta-}) with respect to $M$ 
at fixed  $\Delta_{+}$ and $\Delta_{-}$
and by solving the
resulting
two coupled equations
for the required derivatives $\frac{\partial X}{\partial M}$ and $\frac{\partial D}{\partial M}$.

Applying the conditions for critical points
(Eq.~(\ref{conditioncritical}))
leads to the
following expression for the
surface
$\text{A}_3$
of superfluid
transitions:
\begin{equation}
\label{m zero plane easy}
 T_{s}=\frac{j_{s}}{4}(D+X)\text{.}
\end{equation}
We note that the same relation follows independently from Eq.~(\ref{equiM}) for $H=0$ in the limit $M\rightarrow 0$. The route via Eq.~(\ref{equH}) has, however,
the additional advantage of facilitating also the calculation of tricritical points (see Eqs.~(\ref{conditiontricritical})-(\ref{tcline-constants})).
Furthermore, expanding the right hand side of Eq.~(\ref{equiM}) up to and including the order
$H^3$
leaves the result in Eq.~(\ref{m zero plane easy}) unchanged.

With
$D$ and $X$
given by Eqs.~(\ref{eqx})-(\ref{eqm}) in terms of $\Delta_{+}$, $\Delta_{-}$, and $T$
(note that $H=0$ and that on this surface $M=0$),
Eq.~(\ref{m zero plane easy}) renders $T_s(\Delta_{-},\Delta_{+})$ which
corresponds to a surface in
the space spanned by $(\Delta_{+},\Delta_{-},T)$.

This
surface of
second-order
phase transitions
between the normal fluid $(M=0)$ and
the
superfluid
$(M\neq0)$
ends at the surface of
liquid-vapor
coexistence,
forming a
line of critical 
end points (see the line connecting ce and tce in Fig.~\ref{3dschematic-fig}).
The conditions for tricritical
points
are 
\begin{equation}
\label{conditiontricritical}
 \frac{dH}{dM}|_{\Delta_{+},\Delta_{-},T}=\frac{d^2 H}{dM^2}|_{\Delta_{+},\Delta_{-},T}=\frac{d^3 H}{dM^3}|_{\Delta_{+},\Delta_{-},T}=\frac{d^4 H}{dM^4}|_{\Delta_{+},\Delta_{-},T}=0 \text{, } \frac{d^5 H}{dM^5}|_{\Delta_{+},\Delta_{-},T}>0\text{,}
\end{equation}
with all total
derivatives to be taken
also
at $M=0$, 
which again requires
to consider the partial derivatives of $X$ and $D$ with respect to $M$,
as discussed after Eq.~(\ref{conditioncritical}). The vanishing
of the first four derivatives
leads
to a quadratic equation for $D$
(where Eq.~(\ref{m zero plane easy}) has been used to eliminate
the
dependence on $T$):
\begin{equation}
\label{tcline}
 b_2D^2+b_1D+b_0=0 \text{,}
\end{equation}
where the coefficients
$b_{0,1,2}$
are given in terms of the order parameter $X$ and the coupling constants:
\begin{equation}
\label{tcline-constants}
 \begin{split}
  &b_{0}=X (16 c^2 -(4 j+j_s )(4 k + j_s)) \\
 &\quad \quad +X^2  j_s (4 k + j_s)\text{,}\\
  &b_{1}=2X ( j_s^2-8 c^2 + 8 j k + 2j_s(j+k) )\\ 
 &\quad \quad - (4 c +j_s)^2 + 16 j k \text{,}\\
 &b_{2}=16 c^2 + j_ s (8 c - 4 k + j_s) - 16 j k\text{.}
 \end{split}
\end{equation}
We note that, also here, expanding the right hand side of Eq.~(\ref{equiM}) up
to and including the order $H^2$
does not change the results
in Eqs.~(\ref{tcline})-(\ref{tcline-constants}).

\begin{figure}[t]
\includegraphics[width=95mm]{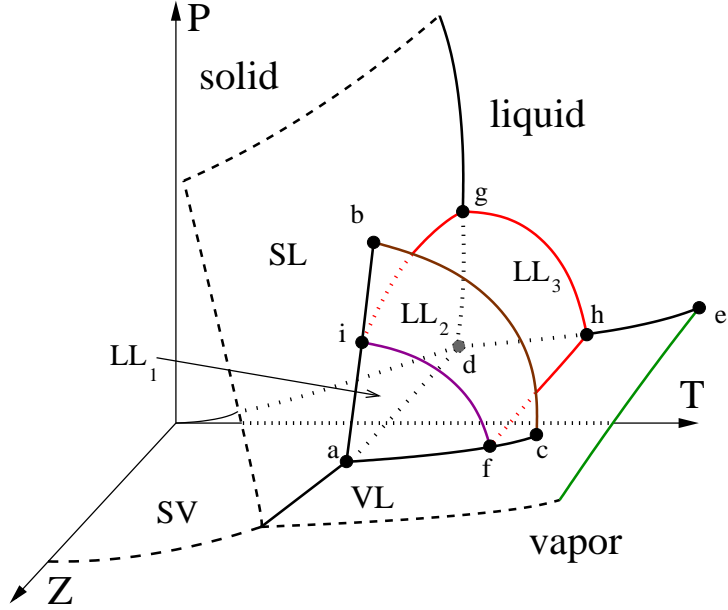}
\centering
\centering
\caption{
\linespread{1.0}
\small{Schematic phase diagram for small
values
of $j_s$. SL, VL, and SV are surfaces of first-order phase transitions
with the same meanings as in
Fig.~\ref{3dA}.
The points denoted as 'g' and 'h' are critical end points of the continuous superfluid transition of the 4-pure fluid
(i.\,e., $Z=0$); g-h is the line of critical points for the continuous superfluid transition of the 4-pure fluid. LL$_3$
is a surface of continuous superfluid transitions bounded by the red lines g-h, h-f, i-g, and the violet line f-i. The triple point of the solid, vapor, and superfluid
phases
of the 4-pure fluid is denoted as 'd'. In the plane $Z=0$ the line to the left of 'd' is the sublimation curve of the 4-pure fluid; d-h-e is the liquid-vapor coexistence line of the 4-pure fluid,
which ends at its critical point 'e'; the extension of the latter to $Z>0$ forms the green line.
The line d-g is the melting curve of the 4-pure
solid
into the superfluid and above 'g' into the normal fluid.
The line d-a is the triple line along which solid, vapor, and superfluid coexist; beyond 'a' this line extends into a triple line along which solid, vapor, and normal fluid coexist. At the quadruple point 'a' solid, vapor, normal fluid, and superfluid coexist. The surface LL$_1$ $\cup$ LL$_2$, which corresponds to the surface LL in Fig.~\ref{3dA},
is the surface of first-order transitions between the 4-rich liquid at the back and the 3-rich liquid in the front; it is bounded by the brown line b-c of critical points which connects the critical end points 'b' and 'c'. The surface
LL$_1$ $\cup$ LL$_2$ of first-order
liquid-liquid
demixing transitions terminates the surface LL$_3$ of continuous superfluid transitions. At this intersection this gives rise to the violet line f-i of critical end points, which themselves end at the
end points 'i' and 'f' of this line of critical end points. At the surface LL$_2$ there are first-order phase transitions between two normal fluids whereas at the surface LL$_1$ there are first-order phase transitions between a normal fluid with high concentration of 3-particles and a superfluid with high concentration of 4-particles.
Accordingly, the superfluid phase forms a dome formed by the plane $Z=0$, SL, VL, LL$_3$, and LL$_1$ with the vertices 'd', 'a', 'f', 'h', 'g', and 'i'.
}}
\label{3dB}
\end{figure}

Accordingly, the solution of Eq.~(\ref{tcline}) yields $D=D_{0}(X)$ which due to Eqs.~(\ref{eqx})-(\ref{eqm}) leads to the relation $D(\Delta_{-},\Delta_{+},T)=D_{0}(X(\Delta_{-},\Delta_{+},T))$. This
turns into a relationship $T(\Delta_{-},\Delta_{+})$ which corresponds to a surface in the
space spanned by
$(\Delta_{+},\Delta_{-},T)$. Simultaneously Eq.~(\ref{m zero plane easy}) has to hold
which also corresponds to a surface in this space. Thus
the tricritical points correspond to the intersection of these two surfaces and thus form a line of tricritical points (TC in Fig.~\ref{3dschematic-fig}).
The condition for the fifth derivative along this line can be checked only numerically. This condition and the fact that $D>|X|$ exclude one of the
two
solutions of Eq.~(\ref{tcline}).
For small values of $j_s$ the model exhibits a superfluid transition in the liquid phase (see Fig.~\ref{3dB}).
In certain parts of the phase diagram this transition is
second order,
in other parts it is
first order.
Thus upon switching on $j_s$ a
new surface LL$_3$
raises above the bottom (i.\,e., VL) of
the phase diagram shown in Fig.~\ref{3dA}
and changes the character of the lower part of the surface LL in Fig.~\ref{3dA}, indicated as LL$_1$, in Fig.~\ref{3dB}.
\\
The surface LL$_3$ of continuous transitions separates the superfluid and the normal fluid
both 4-rich.
The surface LL$_1$ corresponds to first-order phase transitions
between the
4-rich superfluid
and the
3-rich
normal fluid. The surface
LL$_1$ $\cup$ LL$_2$ terminates LL$_3$ at a line f-i of critical end points.

\begin{figure}[t]
\includegraphics[width=95mm]{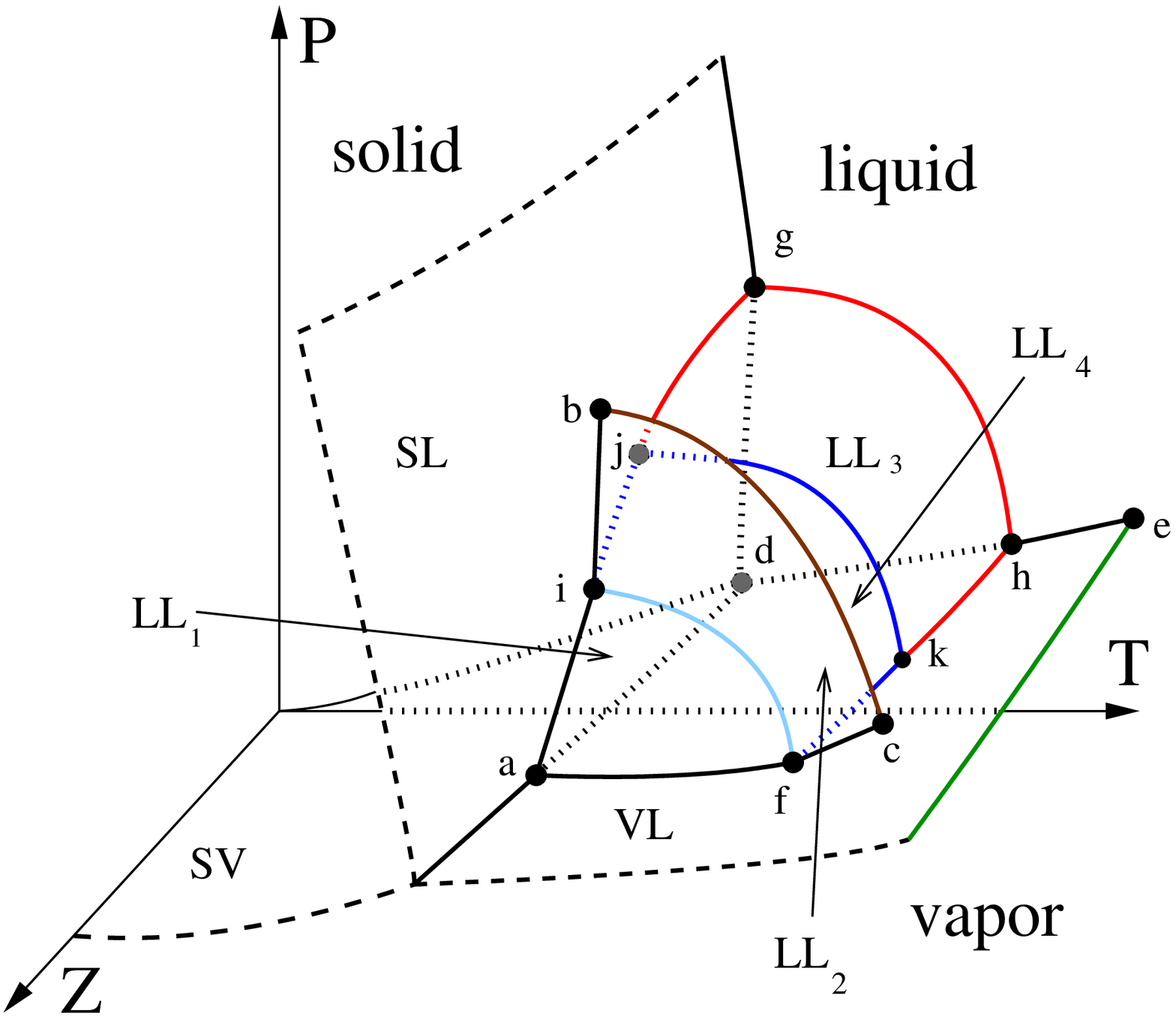}
\centering
\centering
\caption{
\linespread{1.0}
\small{Schematic phase diagram for a value of $j_s$, for which both
critical
(b-c and LL$_3$)
and tricritical
(j-k)
phase transitions between liquids occur.
SL, VL, and SV are the surfaces of first-order phase transitions similar to Fig.~\ref{3dA}.
The superfluid dome is characterized
the vertices
'a', 'f', 'k', 'h', 'g', 'd', 'j', 'i'. Outside this region the liquid is a
normal fluid. LL$_1$ is the surface of first-order phase transitions between the 3-rich
normal fluid and the 4-rich
superfluid,
LL$_2$ (enclosed by the lines i-b, b-c, c-f, f-i) is the surface of first-order
demixing
phase transitions between
3-rich and 4-rich normal fluids,
whereas LL$_3$ is the surface of second-order phase transitions between the normal fluid and the superfluid.
The blue line (j-k) is the line of tricritical points where the surface LL$_3$ connects to the
new
surface LL$_4$
of first-order phase transitions between the normal fluid and the superfluid liquid phases
both being 4-rich.
The surfaces LL$_2$
and
LL$_3$ $\cup$ LL$_4$
meet
at the line of triple points (light blue line i-f).
The brown and the green lines are lines of critical
points;
'a', 'i', and 'f' are quadruple points, 'd' is a triple point, whereas
'b', 'c', 'g', and 'h' are critical end points. The line of triple points (i-f) ends on the surfaces SL and VL at the points 'i' and 'f', respectively.
The points
'j' and 'k' are tricritical end points.
Note that in Fig.~\ref{3dB} the line i-f is a line of critical end points whereas here it is a triple line. This different character motivates their different color code (violet versus light blue). This different character also implies that the
lines a-i-b and a-f-c have a break in slope at 'i' and 'f', respectively, here,
but not in Fig.~\ref{3dB}.
}}
\label{3dC}
\end{figure}

Upon increasing the coupling constant $j_s$
(Fig.~\ref{3dC}),
the model exhibits
as
a new feature a
line
j-k
of tricritical points.
In comparison
with
the phase diagram for weak $j_s$ (Fig.~\ref{3dB}),
a new surface LL$_4$ emerges
(j-k-f-i-j)
which is the surface of first-order phase transitions between
the superfluid and the normal fluid, both
4-rich (Fig.~\ref{3dC}).
The surface LL$_3$ of the second-order phase transitions between the superfluid and the
normal fluid
both 4-rich
meet the surface LL$_4$ at a
line of tricritical points
(dark blue line j-k).
LL$_1$ and LL$_2$
meet LL$_4$ at a triple line
(i-f),
where the superfluid and the
4-rich
normal fluid coexist with the
3-rich
normal fluid.
Thus the increase of $j_s$ changes the character of that part of LL$_3$ in Fig.~\ref{3dB},
which is close to LL$_2$, from
second-order to first-order phase transitions.

\begin{figure}[t]
\includegraphics[width=95mm]{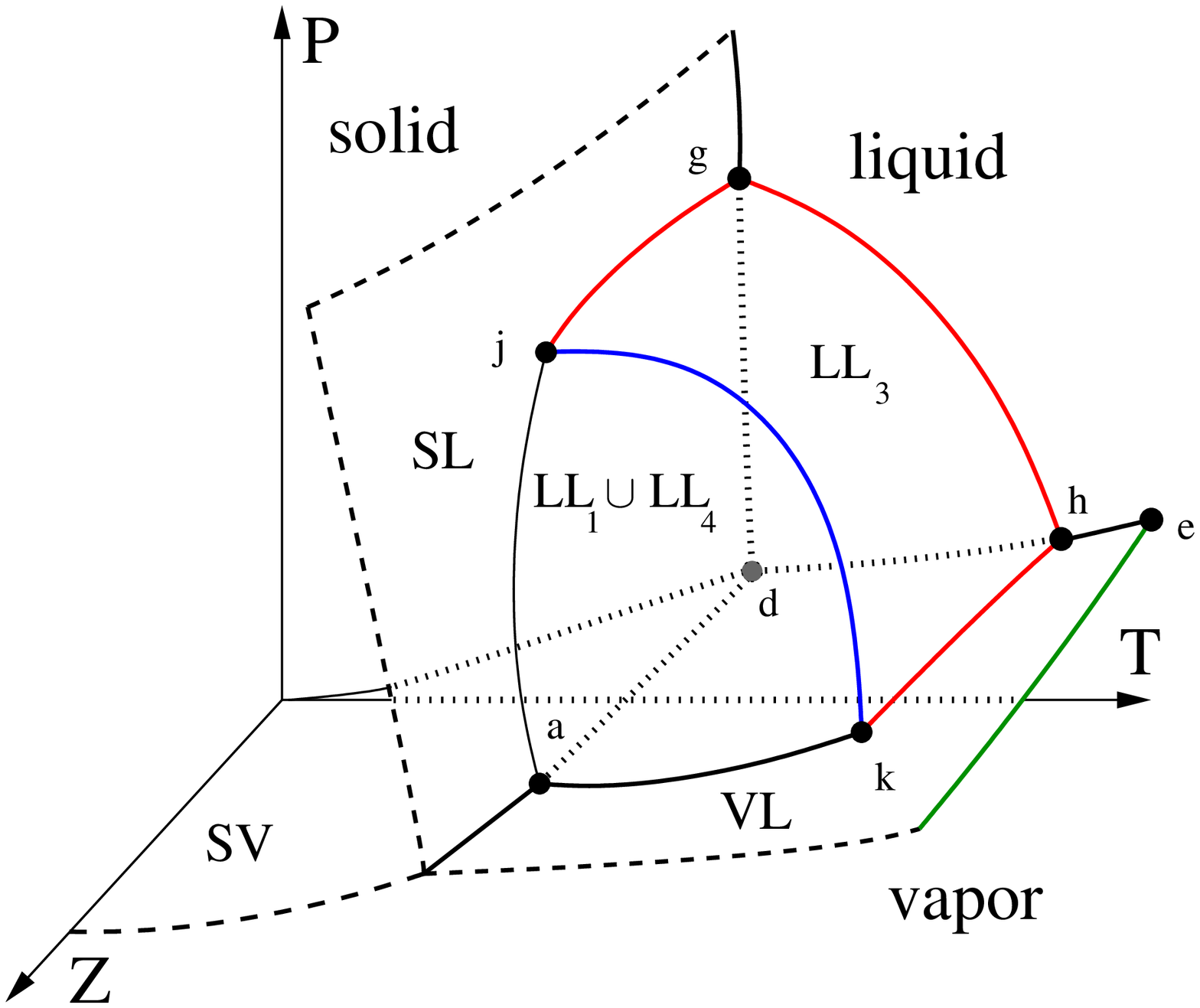}
\centering
\centering
\caption{
\linespread{1.0}
\small{Schematic phase diagram for a value of $j_s$, for which only
a
tricritical line occurs.
In this case,
first-order phase
transitions between liquid phases occur
only between
the superfluid and the normal fluid
so that
the model exhibits only a
(blue)
line of tricritical points (j-k).
SL, VL, and SV are the surfaces of first-order phase transitions
as described in
Fig.~\ref{3dA}. The
superfluid dome is characterized by
the vertices
'a', 'k', 'h', 'g', 'd',
and
'j'. Outside this region the liquid is a normal fluid.
The surface LL$_2$ from Fig.~\ref{3dC} does
not exist anymore
and the transitions
between
liquid phases are either second-order phase transitions between the normal fluid and the superfluid mixed liquid (LL$_3$), or
first-order phase transitions between the normal fluid and the superfluid liquid (LL$_1\cup$LL$_4$). The surfaces LL$_3$ and LL$_1\cup$LL$_4$ meet at the line
j-k
of tricritical points (blue line).
The points
'g' and 'h' are critical end points, whereas 'j' and 'k' are tricritical end points.
The point
'a' is a quadruple point and 'd' is a triple point.}}
\label{3dD}
\end{figure}

If
the coupling constant $j_s$ is
increased further (Fig.~\ref{3dD}),
first-order phase
transitions
between
liquid phases
occur
only between
the superfluid and the normal fluid
phase.
There are no longer first-order demixing transitions between two normal fluids.
Thus, upon increasing $j_s$, the surface LL$_3\cup$LL$_4$ moves up
(i.e.,
towards higher $P$ and $T$) so that accordingly the line i-f
also moves up towards the line b-c. This implies that LL$_2$ shrinks and the wedge between the lines i-b and i-j becomes shorter. Finally LL$_2$ and b-c disappear and
LL$_1$ and LL$_4$ become a single surface of first-order transitions between 4-rich superfluid and 3-rich normal liquid; this implies that the line i-f disappears, too.
Accordingly, the phase diagram is left with only a (blue) line of tricritical points k-j.
This
topology
of the phase diagram
is shown in Fig.~\ref{3dD}.
In this case the liquid-liquid phase transitions are either second-order phase transitions
on LL$_3$
between the normal fluid and the superfluid mixed liquid, or
first-order phase transitions
on LL$_4\cup$LL$_1$
between the normal fluid and the superfluid liquid.

As discussed in the introduction, in the case of actual
$\Ht$ - $\Hf$
mixtures
the solid phase is formed only at
high pressures, whereas for sufficiently low pressures
the superfluid reaches down to $T=0.$
In order to obtain this topology from that of
Fig.~\ref{3dD},
by fiat one has to pull up
and tilt the surface SL and to shift the superfluid dome down to $T=0$ so that the surface SV disappears.
This
transforms the phase diagram in Fig.~\ref{3dD} to the one shown in Fig.~\ref{3dschematic-fig}
such that g = ce$^+$, h = ce, e = c, LL$_3$ = A$_3$, j-k = TC, j = tce$^+$, k = tce, and
LL$_1\cup$LL$_4$ = A$_4$.
In this sense the bulk phase diagram shown in Fig.~\ref{3dD} is supposed to mimic
the one of
the actual $\Ht$- $\Hf$ mixtures.

\begin{figure}[t]
\includegraphics[width=85mm]{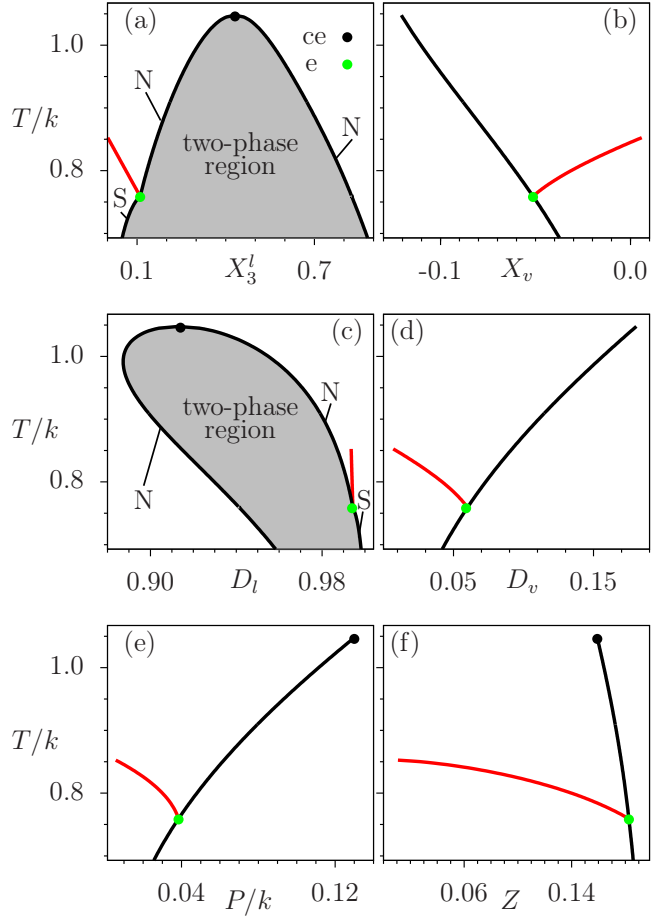}
\centering
\centering
\caption{
\linespread{1.0}
\small{
Phase diagram
for the coupling constants $(c/k, j/k, j_s/k)=(1, 5.714, 1.717)$ and $H=0$
corresponding to Fig.~\ref{3dB}.
Along the triple line
a-f in Fig.~\ref{3dB}
the figures show
the coexistence between vapor, normal fluid, and superfluid
(a) in the $(X^{l}_3$, $T)$ plane, with $X^{l}_3=\langle N_3\rangle/\mathcal{N}$ corresponding to the
number density of
3-particles in the liquid phase, (b) in the $(X_{v}$, $T)$ plane at
coexistence with the two liquid phases, where $X_v=(\langle N_4\rangle_v-\langle N_3\rangle_v)/\mathcal{N}=\langle X_4^v\rangle-\langle X_3^v\rangle$ in the vapor phase, (c) in the $(D_{l}$, $T)$ plane,
where $D_l=(\langle N_4\rangle_l+\langle N_3\rangle_l)/\mathcal{N}$
is the total number density
in the liquid phase, and (d) in the $(D_{v}$, $T)$ plane,
where $D_v=(\langle N_4\rangle_v+\langle N_3\rangle_v)/\mathcal{N}$
is the total number density
in the vapor phase. The indices
'$l$' and '$v$' refer to the values of the order parameters in the liquid and the vapor
phase,
respectively.
Panels (e) and (f) show the
temperature dependence of $P$ and $Z$ along the triple line a-c Fig.~\ref{3dB} (black)
and along the red line f-h near 'f' in Fig.~\ref{3dB}.
The red line is the line
of second-order
transitions between the normal fluid and superfluid at coexistence with vapor,
which ends at the demixing curve at the
green
critical end point 'e' (i.\,e., 'f' in Fig.~\ref{3dB}). At 'e' the liquid state $(X^{\text{e}}_l, D^{\text{e}}_l, M^{\text{e}}_l)=(0.773, 0.994, 0)$
coexists with
the
vapor
state
$(X^{\text{e}}_v, D^{\text{e}}_v, M^{\text{e}}_v)=(-0.051, 0.059, 0)$ at $T_{\text{e}}/k=0.758$. N and S denote normal
fluid
and superfluid, respectively.
The transitions between the vapor and the liquid phases are
always
first order. The points 'ce' and 'e' here correspond to the points 'c' and 'f' in Fig.~\ref{3dB}.
}}
\label{t2}
\end{figure}

\begin{figure}[t]
\includegraphics[width=85mm]{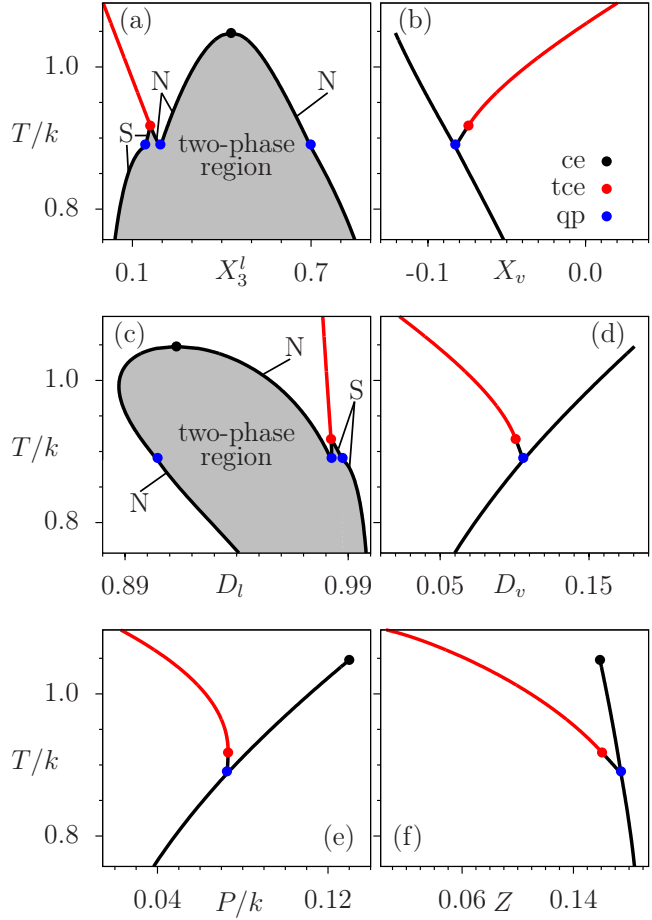}
\centering
\centering
\caption{
\linespread{1.0}
\small{
Phase diagram
for the coupling constants $(c/k, j/k, j_s/k)=(1, 5.714, 2.231)$ and $H=0$
which corresponds to Fig.~\ref{3dC}.
Along the triple
lines a-f and f-c in Fig.~\ref{3dC}
the figures show the first-order demixing transitions
of
the liquid phase at coexistence with the vapor phase
(a) in the $(X^{l}_3$, $T)$ plane, with
$X^{l}_3=\langle N_3\rangle_l/\mathcal{N}$
as the number density of
3-particles in the liquid phase, (b) in the $(X_{v}$, $T)$ plane at
coexistence
of the vapor
with the two liquid phases, where $X_v=(\langle N_4\rangle_v-\langle N_3\rangle_v)/\mathcal{N}$ in the vapor phase, (c) in the $(D_{l}$, $T)$ plane,
where $D_l=(\langle N_4\rangle_l+\langle N_3\rangle_l)/\mathcal{N}$ in the liquid phase, and (d) in the $(D_{v}$, $T)$ plane,
where $D_v=(\langle N_4\rangle_v+\langle N_3\rangle_v)/\mathcal{N}$  in the vapor phase. The indices
'$l$' and '$v$' refer to the values of the order parameters in the liquid and the vapor
phase, respectively. Panels (e) and (f) show the
temperature dependence of $P$ and $Z$ along the triple lines a-f, f-c, and
f-k, in Fig.~\ref{3dC}.
The red line corresponds to second-order
phase transitions between normal fluids and superfluids (line k-h in Fig.~\ref{3dC}).
At tce, the liquid state $(X^{\text{tce}}_l, D^{\text{tce}}_l, M^{\text{tce}}_l)=(0.662, 0.982, 0)$ coexists with the vapor
state $(X^{\text{tce}}_v, D^{\text{tce}}_v, M^{\text{tce}}_v)=(-0.074, 0.100, 0)$ at $T_{\text{tce}}/k=0.917$.
The point
ce remains as in the case $j_s=0$. N and S denote normal
liquid
and superfluid, respectively. At the quadruple point 'qp'
the four coexisting states at $T_{\text{qp}}/k=0.887$ are two normal liquids $(X^{\text{qp}}_l,D^{\text{qp}}_l,M^{\text{qp}}_l)=\{(-0.493, 0.904, 0),(0.594, 0.982, 0)\}$, a superfluid
$(X^{\text{qp}}_l, D^{\text{qp}}_l, M^{\text{qp}}_l)=(0.701, 0.987, 0.281)$, and
the vapor state $(X^{\text{qp}}_v, D^{\text{qp}}_v, M^{\text{qp}}_v)=(-0.082, 0.105, 0)$.
The points 'ce', 'tce', and 'qp' here correspond to the points 'c', 'k', and 'f', respectively in Fig.~\ref{3dC}.
In (b), (d), (e), and (f), the long black coexistence curves are expected to exhibit a break in slope at 'qp'; on the present scales this is not visible.
}}
\label{t3}
\end{figure}

\begin{figure}[t]
\includegraphics[width=85mm]{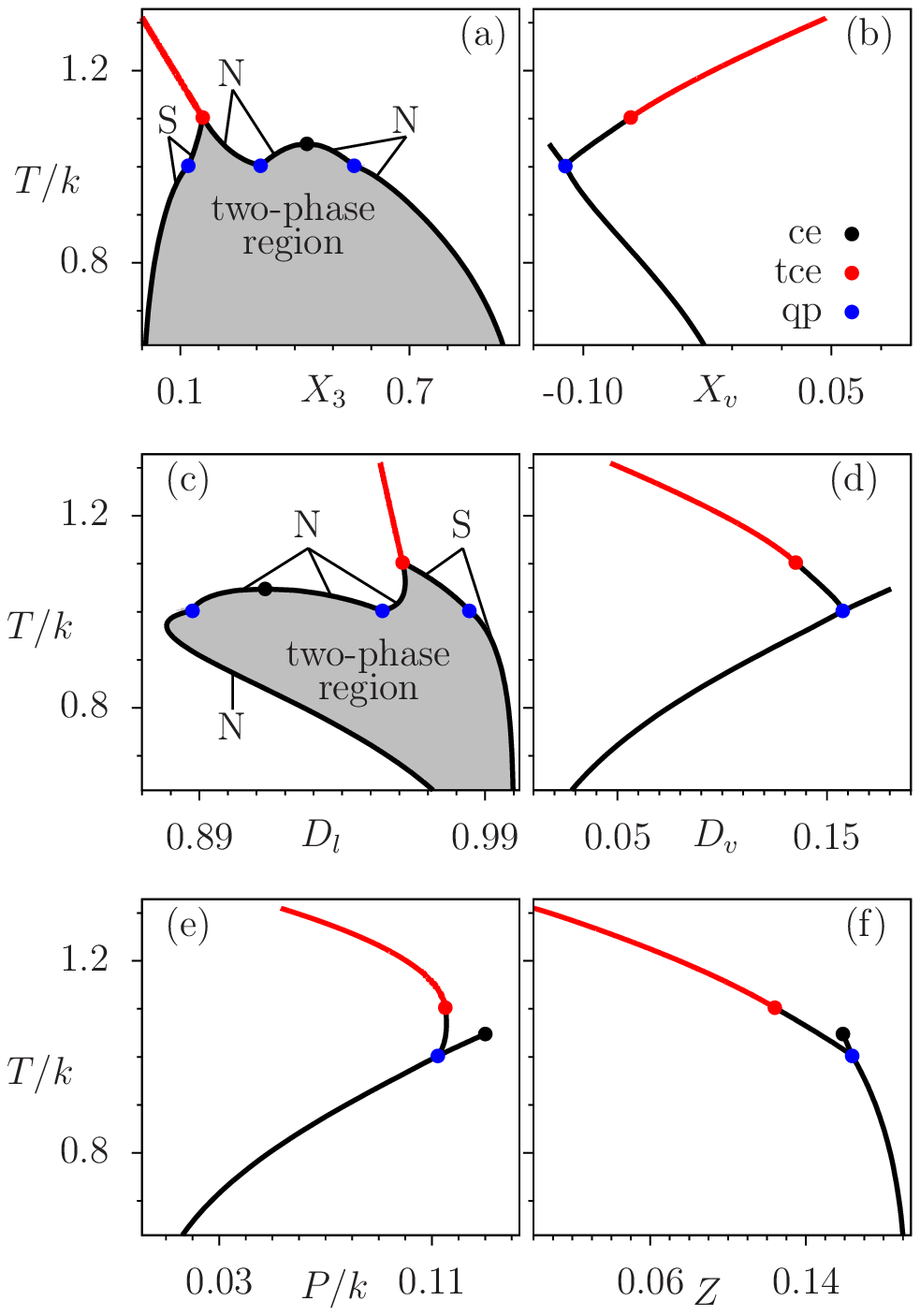}
\centering
\centering
\caption{
\linespread{1.0}
\small{The same as in Fig.~\ref{t3} but for the coupling constants $(c/k, j/k, j_s/k)=(1, 5.714, 2.747)$ and $H=0$.
At tce the order parameters of the liquid and the vapor phases are $(X^{\text{tce}}_l, D^{\text{tce}}_l, M^{\text{tce}}_l)=(0.644, 0.961, 0)$ and $(X^{\text{tce}}_v, D^{\text{tce}}_v, M^{\text{tce}}_v)=(-0.071, 0.135, 0)$,
respectively,
and
$T_{\text{tce}}/k=1.101$.
Again,
ce remains as in the case of $j_s=0$. N and S denote normal
liquid
and superfluid, respectively.
The four coexisting states at
the quadruple point
'qp' are given by $(X^{\text{qp}}, D^{\text{qp}}, M^{\text{qp}})=(-0.222, 0.888, 0), (0.334, 0.954, 0), (0.742, 0.984, 0.456), \text{and } (0.110, 0.157, 0)$
at the temperature $T_{\text{qp}}/k=1.001$.
The long black lines in (b) and (d) and the ones in (e) and (f) ending at 'ce' are expected to exhibit a break in slope at 'qp'; on the present scales this is not visible.
}}
\label{t4}
\end{figure}

\begin{figure}[t]
\includegraphics[width=85mm]{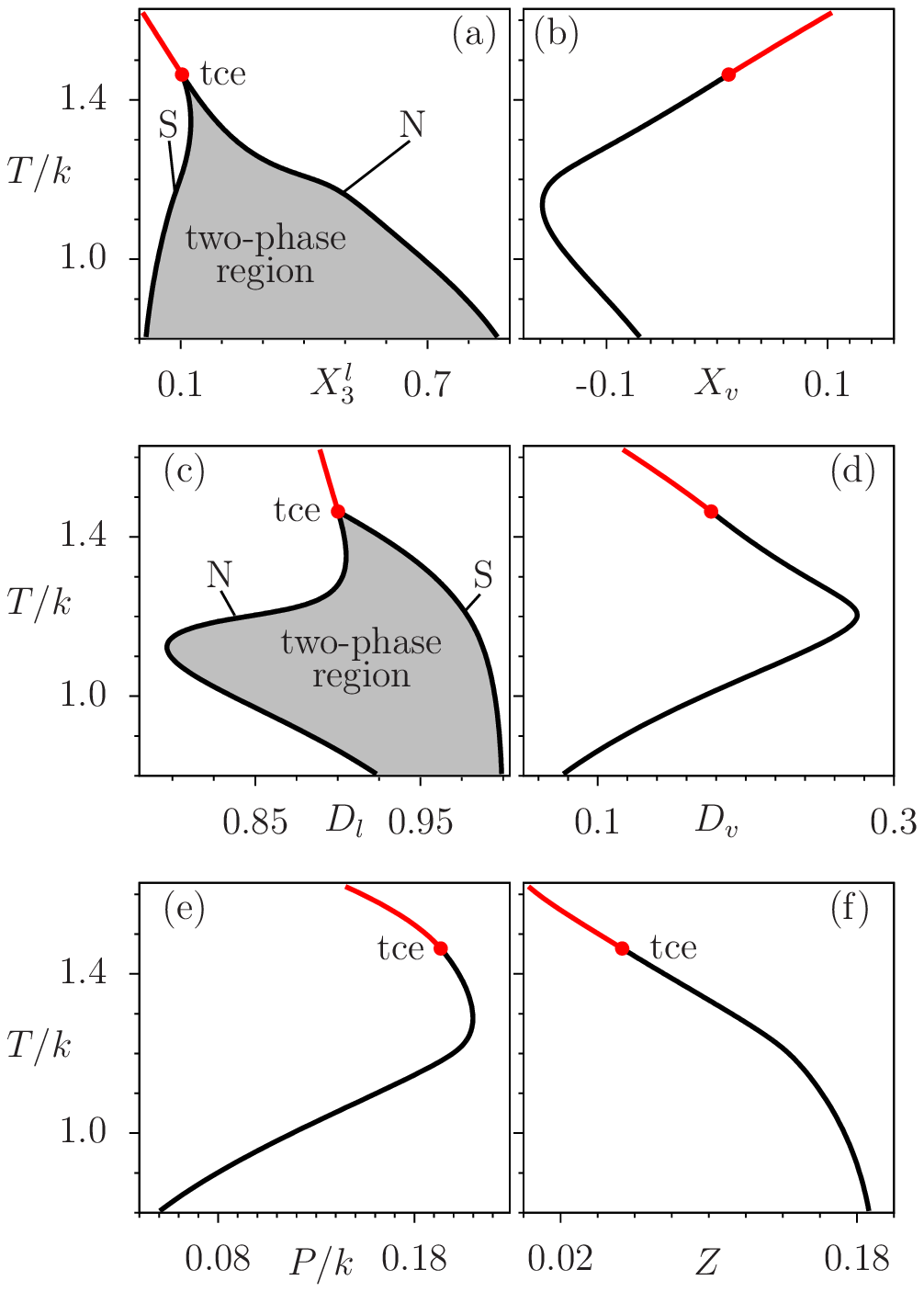}
\centering
\centering
\caption{
\linespread{1.0}
\small{Phase diagrams for the coupling constants $(c/k, j/k, j_s/k)=(1, 5.714, 3.674)$ and $H=0$,
corresponding to Fig.~\ref{3dD}.
The panels show the same as in Figs.~\ref{t3} and~\ref{t4};
however the critical end point 'ce' has disappeared.
At tce
the
liquid state $(X^{\text{tce}}_l, D^{\text{tce}}_l, M^{\text{tce}}_l)=(0.693, 0.900, 0)$ coexists with
the
vapor
state}
$(X^{\text{tce}}_v, D^{\text{tce}}_v, M^{\text{tce}}_v)=(0.010, 0.176, 0)$ at $T_{\text{tce}}/k=1.462$.
For
$T>T_{\text{tce}}$ there is only
a second-order phase transition from
a
normal mixed liquid to
a
superfluid.
For $T<T_{\text{tce}}$ the phase
transitions between the normal fluid and the superfluid
are first order.
N and S denote normal
liquids
and
superfluids,
respectively.}
\label{t5}
\end{figure}

The demixing transitions at coexistence with the vapor phase for
various
sets of the coupling constants are shown in
Figs.~\ref{t2}-\ref{t5}. In these figures the values of $c/k$ and $j/k$ are the same; only the value of $j_s$ is changed.
For the choice of coupling constants
$(c/k, j/k, j_s/k)=(1, 5.714, 1.717)$ (see Fig.~\ref{t2}), the phase diagram
exhibits the topology of
the schematic phase diagram
shown
in Fig.~\ref{3dB}. The red line in
Fig.~\ref{t2}(a)
provides the temperature dependence of $X_3$ along the red line in Fig.~\ref{3dB} emanating from 'f' towards 'h'.
The green point 'e'
in Fig.~\ref{t2}
corresponds to the point 'f' in Fig.~\ref{3dB} and the black point in Fig.~\ref{t2} corresponds to the point 'c'
in Fig.~\ref{3dB}.
Because
in Fig.~\ref{3dB} the red line h-f is
a
line of continuous phase transitions right up to the point
'f',
the line a-f-c does not exhibit a break in slope at 'f'.
Below e, the liquid transitions
are first-order transitions between the normal fluid and the superfluid liquid, whereas above e the demixing curve remains the same as in the case of $j_s=0$
(see Fig.~\ref{3dA}).

As discussed in Fig.~\ref{3dC}, for
even
larger values of $j_s$, both
continuous and first-order superfluid transitions occur, giving rise to
the
occurrence of a line of
tricritical points.
For the choice of coupling constants $(c/k, j/k, j_s/k)=(1, 5.714, 2.231)$ and $(c/k, j/k, j_s/k)=(1, 5.714, 2.747)$
Figs.~\ref{t3} and \ref{t4},
respectively,
show the liquid-liquid transitions at coexistence with the vapor phase
for such a topology
of the phase diagram.
In both figures one
finds
two types of first-order liquid-liquid transitions. One between two
normal
liquids,
which occur between ce and qp,
and another one between the normal liquid phase and the superfluid liquid phase, which
occur
below tce.
The points 'ce', 'tce', and 'qp' in Figs.~\ref{t3} and \ref{t4} correspond to the points
'c', 'k', and 'f', respectively, in Fig.~\ref{3dC}. The transitions between the two normal
liquids
correspond to the line f-c in
Fig.~\ref{3dC},
the transitions between the normal
liquid
and the superfluid correspond to the line
a-f in Fig.~\ref{3dC}, the small two phase region between 'tce' and 'qp' corresponds to the line f-k in Fig.~\ref{3dC},
and the red line above 'tce' corresponds to the red line emanating from 'k' towards 'h' in Fig.~\ref{3dC}.
In Fig.~\ref{3dC} the triple lines a-f and k-f
merge at the quadrupole point
'qp' = 'f',
where four
phases
coexist: two normal
liquids,
the superfluid, and the vapor phase. Below 'qp', the liquid-liquid transitions
at coexistence with the vapor phase are first-order transitions between the normal fluid and the superfluid.
Upon increasing $j_s$ the tricritical end point
tce = k
is pulled towards higher temperatures (compare Figs.~\ref{t3} and \ref{t4}).

In order to
obtain
phase diagrams
with the topology
illustrated in
Fig.~\ref{3dD}, one has to
choose
the coupling constants such that the demixing transitions at coexistence with the vapor phase
occur
only between the normal fluid and the superfluid.
This means that in Fig.~\ref{3dC} the line f-c has to shrink to zero which implies that the critical point 'c' coincides with the quadruple point 'f'. Within Fig.~\ref{t4}(a) this means that tce (= k in Fig.~\ref{3dC})
has to be pulled up to higher temperatures such that the demixing critical end point ce (= c in Fig.~\ref{3dC})
slides below the quadruple qp (= f in Fig.~\ref{3dC}) so that the demixing phase transition between two normal fluids
becomes an unstable one within the
two-phase
region of the superfluid and the mixed normal fluid (see Fig.~\ref{t5}(a)).
For the coupling constants $(c/k, j/k)=(1, 5.714)$ this is fulfilled, provided that $j_s/k>2.96$.
For the coupling constants $(c/k, j/k, j_s/k)=(1, 5.714, 2.96)$ at $T_{\text{ce}}/k=1.047$ only three thermodynamic states coexist:
the critical state $(X_{\text{ce}},D_{\text{ce}},M_{\text{ce}})=(0.050, 0.913, 0)$, the vapor phase,
and a superfluid state $(X_{\text{s}}, D_{\text{s}}, M_{\text{s}})=(0.756, 0.983, 0.497)$.
Accordingly, for coupling constants $(c/k=1, j/k=5.714, j_s/k>2.96)$ one obtains the type of phase diagram shown in Figs.~\ref{3dD} and~\ref{t5}.

As can be inferred from Fig.~\ref{t5}(f), upon increasing the temperature, the line of second-order phase transitions to the superfluid phase (red line) approaches
the plane $Z=0$,
where the liquid becomes pure $\Hf$. In order to
explore the phase diagram in the plane $Z=0$,
in Eqs.~(\ref{eqx}) to (\ref{EqR})
we have to take the limit
$\mu_{3}\rightarrow -\infty$. 
In this limit
$\Delta_{-}\rightarrow +\infty$ and $\Delta_{+}\rightarrow -\infty$ so that
$W$ and $R$
turn into

\begin{equation}
\lim_{\mu_{3}\rightarrow -\infty} W(\Delta_{-},\Delta_{+},H,T)=0, \quad 
\end{equation}
and
due to $\Delta_{-}+\Delta_{+}=\mu_{4}$
\begin{equation}
\lim_{\mu_{3}\rightarrow -\infty} R(\Delta_{-},\Delta_{+},H,T)=e^{\beta((c+k)X+(j+c)D+\mu_4)}. \quad 
\end{equation}
Since $W_{\mu_{3}\rightarrow -\infty}=0$, due to Eqs.~(\ref{eqx}) and~(\ref{eqd})
one has
$X_{\mu_{3}\rightarrow -\infty}=D_{\mu_{3}\rightarrow -\infty}$, where
$D_{\mu_{3}\rightarrow -\infty}$ is given by
\begin{equation}
\label{d=x}
\begin{split}
D_{\mu_{3}\rightarrow -\infty} &=\lim_{\mu_{3}\rightarrow -\infty} \frac{R(\Delta_{-},\Delta_{+},H,T)I_0(j_{s}M/T+H/T)}{1+R(\Delta_{-},\Delta_{+},H,T)I_0(j_{s}M/T+H/T)}\\
                               &=\frac{e^{\beta((j+k+2c)D+\mu_4)}I_0(j_{s}M/T+H/T)}{1+e^{\beta((j+k+2c)D+\mu_4)}I_0(j_{s}M/T+H/T)}, \quad 
\end{split}
\end{equation}
where,
due to $X_{\mu_{3}\rightarrow -\infty}=D_{\mu_{3}\rightarrow -\infty}$,
in $R_{\mu_{3}\rightarrow -\infty}$ we have replaced $X$ by $D$.

In this limit Eq.~(\ref{equiM}) reduces to
\begin{equation}
\frac{D}{M}=\frac{I_0(j_{s}M/T+H/T)}{I_1(j_{s}M/T+H/T)}
\end{equation}

and
the equilibrium free energy
(Eq.~(\ref{free energy}))
reduces to 
\begin{equation}
\phi (\mu_{3}\rightarrow -\infty,\mu_4,H,T)=\frac{k+j+2c}{2}D^2+\frac{j_s}{2}M^2+T \ln (1-D).
\end{equation}

In this limit the
temperature of the superfluid transition is given by
\begin{equation}
T_{\text{s}}=\frac{j_s}{2}D,
\end{equation}

and $\mu_4$
follows
from Eq.~(\ref{d=x}):
\begin{equation}
\label{mu4EQ}
\mu_4(H,T)=T\ln \frac{D}{1-D}-(j+2c+k)D-T\ln I_0(j_{s}M/T+H/T).
\end{equation}

\begin{figure}[t]
\includegraphics[width=85mm]{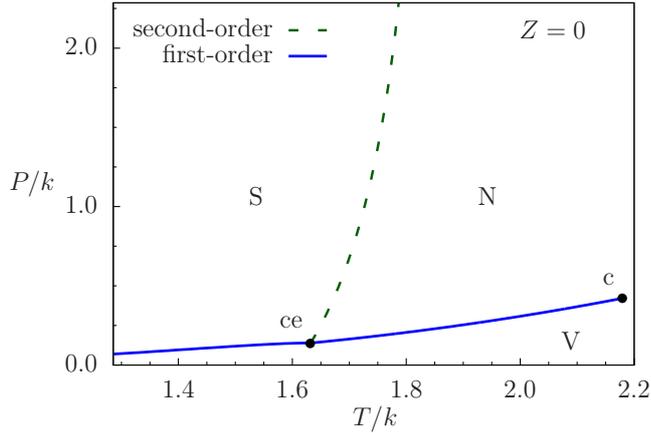}
\centering
\centering
\caption{
\linespread{1.0}
\small{The
$(T, P)$
phase diagram for the coupling constants $(c/k, j/k, j_s/k)=(1, 5.714, 3.674)$ and $H=0$ for pure $\Hf$,
i.\,e.,
$Z=0$.
The dashed green line shows the $\lambda$-line of second-order phase transitions
between normal
liquids
and superfluids.
N, S, and V denote
the normal
liquid,
the superfluid, and the
vapor phase, respectively. The blue line
of first-order
liquid-vapor transitions terminates
the $\lambda$-line
at the critical
end point 'ce' and ends at the critical point 'c' of
the liquid-vapor
coexistence
line.
( We have been unable to find a set of coupling constants for which the dashed $\lambda$-line of second-order phase transitions exhibits a negative
slope as it is the case for actual $\Hf$.)
}}
\label{ptZERO}
\end{figure}

For pure $\Hf$, i.\,e., for $Z=0$ and for the choice of the coupling constants $(c/k, j/k, j_s/k)=(1, 5.714, 3.674)$,
the phase diagram in the $(T$, $P)$ plane  is shown in Fig.~\ref{ptZERO}.
The dashed green line shows the $\lambda$-line of second-order phase transitions
between normal liquids and superfluids.
This line
is terminated by
the line of first-order liquid-vapor phase transitions (blue line) at the critical end point ce.
The line of first-order liquid-vapor phase transitions
ends at the critical point c.
For high
pressures the system becomes solid,
(see Fig.~\ref{he4}) which, however, is not captured by the present model.
Along the line of first-order liquid-vapor transitions
($T>T_{\text{ce}}$, blue line in Fig.~\ref{ptZERO}),
the difference between
the
number densities of the liquid and the vapor
phase decreases
upon increasing the temperature and
vanishes at $T=T_{\text{c}}$.
Accordingly, the
two phases merge into a single phase at
the critical point
'c'
given by
\begin{equation}
\label{critical-liquid-vapor}
\frac{\text{d}\mu_{4}}{\text{d}D}|_{T}=\frac{\text{d}^2 \mu_{4}}{\text{d}D^2}|_{T}=0 \text{, } \quad \frac{\text{d}^3 \mu_{4}}{\text{d}D^3}|_{T}>0\text{,}
\end{equation}
where $\mu_4$ is given by Eq.~(\ref{mu4EQ}). These conditions reduce to
(note that $I_0(0)=1$)
\begin{equation}
D_{\text{c}}=0.5 \text{,} \quad T_{\text{c}}=0.25 (2 c + j + k)\text{.}
\end{equation}

For nonzero values of $Z$, i.e., in the presence of $\Ht$ atoms, the critical points of the
phase transitions between vapor and normal liquids
($M=0$) are given by (see  Eqs.~(\ref{coupling}) and~(\ref{eqdelta+}))
\begin{equation}
\label{criticalCond-L-V}
\frac{\mathrm{d}\Delta_{+}}{\mathrm{d}D}|_{\Delta_{-},T}=\frac{\mathrm{d}^2\Delta_{+}}{\mathrm{d}D^2}|_{\Delta_{-},T}=0\text{,} \quad \frac{\mathrm{d}^3\Delta_{+}}{\mathrm{d}D^3}|_{\Delta_{-},T}>0\text{,}
\end{equation}
where
in Eq.~(\ref{eqdelta+}) also
the partial derivatives of $X$ with respect to $D$ must be
taken into account.
%
%
%
%
%
\begin{figure}[t]
\includegraphics[width=120mm,clip=true,trim=22 26 35 24]{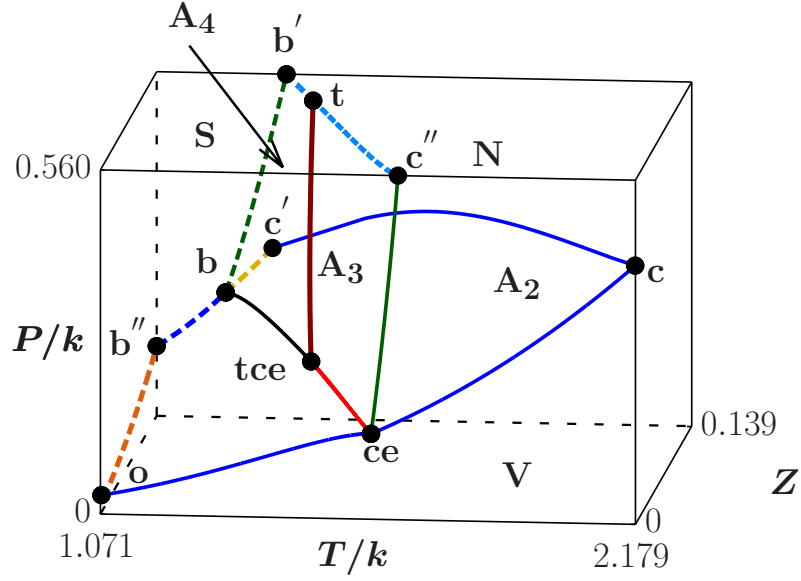}
\centering
\centering
\caption{
\linespread{1.0}
\small{
Numerical results for the
fluid parts of the phase diagram for the choice of the coupling constants $(c/k, j/k, j_s/k)=(1, 5.714, 3.674)$ and $H=0$ in $(P$, $Z$, $T)$ space.
The points o, ce, c, and c$^{''}$
lie
in the zero fugacity plane ($Z=0$) and the points c$^{''}$, t, and b$^{'}$
lie in
the constant pressure plane $P/k=0.560$.
The points
o and
b$^{''}$
have the same temperature $T/k=1.071$,
while b, b$^{'}$, and c$^{'}$ share the same value of
the
fugacity $Z=0.139$.
The surface
(o-ce-tce-b-b$^{''}$-o)
corresponds to
first-order phase transitions between the vapor
phase
(V) and the superfluid
phase
(S),
whereas
(ce-c-c$^{'}$-b-tce-ce)
is the surface of first-order phase transitions between the vapor
phase
and the normal liquid
phase
(N);
their union corresponds to A$_2$ in Fig.~\ref{3dschematic-fig}.
The surface
(b-tce-t-b$^{'}$-b)
is the surface of first-order phase transitions between the superfluid and the normal
liquid phase
corresponding to A$_4$ in Fig.~\ref{3dschematic-fig}
and
(tce-ce-c$^{''}$-t-tce)
is the surface of second-order phase transitions between the superfluid and the normal
liquid phase
corresponding to A$_3$ in Fig.~\ref{3dschematic-fig}.
The black line b-tce and the
light
red line tce-ce are the lines of first- and second-order liquid-liquid transitions at coexistence with the vapor phase, respectively,
which meet at the tricritical end point tce.
The solid blue line
c-c$^{'}$
is the line of critical points of the liquid-vapor phase transitions and the
dark
red curve (tce-t) is the line of tricritical points. The lowest pressure is $p/k=0$
whereas the highest temperature is $T/k=2.179$.
The line o-b$^{''}$ is the intersection of A$_2$ and the plane $T/k=1.071$, the line c$^{'}$-b-b$^{''}$ is the intersection of A$_2$ and the plane $Z=0.139$;
the line b-b$^{'}$ is the intersection of A$_4$ with the plane $Z=0.139$; the line b$^{'}$-t and t-c$^{''}$ are the intersection of A$_4$ and A$_3$, respectively, with the plane $P/k=0.560$.
We note that at 'ce' the line o-ce-c does not exhibit a break in slope (see Fig.~\ref{ptZERO}).
}}
\label{3d}
\end{figure}
Having determined various features of the phase diagram of
the present model for a set of coupling constants for which
the topology of the phase diagram is that of the experimental one, we can illustrate quantitatively
the phase diagram in the $(P$, $Z$, $T)$ space.
%
%
%
%
%
%
%
%
%
%
%
%
%
%
%
%
%
%
%
%
%
%
%
%
%
The phase diagram,
which --  for a suitable set of coupling constants --
resembles the schematic phase diagram proposed in Ref.~\cite{1922}
and exhibits all relevant fluid phases,
is given in Fig.~\ref{3d} (compare Fig.~\ref{3dschematic-fig}).
%
Accordingly, Fig.~\ref{3d} shows where the vapor phase (V), the normal
liquid
phase (N),
and the superfluid phase are thermodynamically stable and where first- or second-order phase transitions among each other occur.
The transitions between the vapor and the liquid phases are given by the two surfaces
o-ce-tce-b-b$^{''}$-o
and
ce-c-c$^{'}$-b-tce-ce (the union of which corresponds to $\text{A}_2$ in Fig.~\ref{3dschematic-fig}),
while the
loci of the phase
transitions between the superfluid and the normal fluid
form
the two surfaces
b-tce-t-b$^{'}$-b
and
tce-ce-c$^{''}$-t-tce
which
in Fig.~\ref{3dschematic-fig},
correspond to $\text{A}_4$ and $\text{A}_3$, respectively.

The points o, ce, c, and c$^{''}$
lie
in the zero fugacity plane ($Z=0$)
whereas
b$^{'}$, t, and c$^{''}$ lie in
the plane of
constant pressure $P/k=0.560$.
The points b$^{''}$ and o
are located in the plane of constant
temperature $T/k=1.071$, while b, b$^{'}$, b$^{''}$, and c$^{'}$ share the same
value of fugacity $Z=0.139$. The black
line
b-tce and the
light
red
line
tce-ce
indicate
first- and second-order liquid-liquid phase transitions,
respectively,
at coexistence with the vapor phase. These two
lines
are connected
at
the tricritical
end point
tce. The
dark
red solid line (tce-t) connects the surfaces
A$_4$ and A$_3$
of first- and second-order liquid-liquid phase transitions
((b-b$^{'}$-t-tce-b) and (t-tce-ce-c$^{''}$-t)),
respectively.
The coexisting states along
the two lines
(b-tce, $T<T_{\text{tce}}$) and (tce-ce, $T>T_{\text{tce}}$)
are the ones shown in Fig.~\ref{t5}. The solid blue line (c-c$^{'}$) is the line of critical points of the liquid-vapor phase transitions and the
dark
red curve (tce-t) is the line of tricritical points
with the tricritical end point tce.

By moving along the line b-b$^{''}$ towards b$^{''}$ the number density
in
the liquid phase increases. This implies that the larger the number density of the liquid phase at b
is,
the shorter
is
the line b-b$^{''}$ (note that $D<1$). This means that, by lowering the temperature along the line of first-order liquid-liquid phase transitions at coexistence with the vapor phase (tce-b),
the point b shifts towards the point b$^{''}$.

\begin{figure}[t]
\includegraphics[width=85mm]{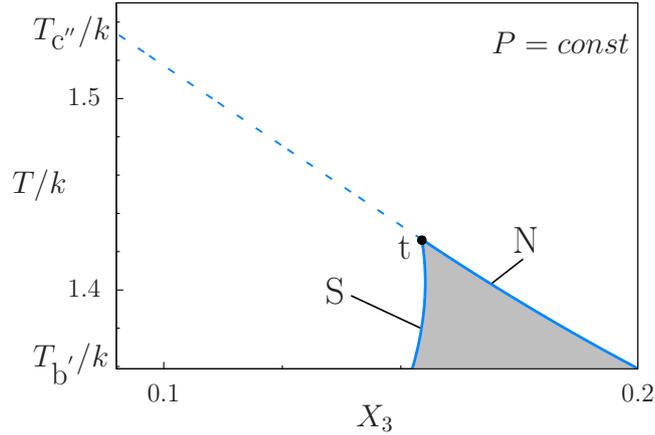}
\centering
\centering
\caption{
\linespread{1.0}
\small{The liquid-liquid phase transitions at fixed pressure $P/k=0.560$ in the $(X_3$, $T)$ plane for the choice of the coupling constants $(c/k, j/k, j_s/k)=(1, 5.714, 3.674)$ and $H=0$.
The figure provides the temperature dependence of $X_3$ along the line b$^{'}$-t-c$^{''}$ in Fig.~\ref{3d}. For $T_{\text{t}}<T<T_{\text{c$^{''}$}}$ the light blue dashed line
represents continuous phase transitions whereas for $T_{\text{b$^{'}$}}<T<T_{\text{t}}$ the lines indicate the coexisting superfluid (S) and normal liquid (N)
states at first-order phase transitions.
The two-phase region is shaded in grey. The point t corresponds to a tricritical point.}}
\label{xtLIQUID}
\end{figure}

The liquid-liquid phase transitions at constant pressure are given by the
curve
b$^{'}$-t-c$^{''}$.
The curve (b$^{'}$-t) is
a
line of first-order liquid-liquid phase transitions at constant pressure, which is connected to
the line of second-order liquid transitions (t-c$^{''}$)
at
the tricritical point t. The coexisting states along these two lines
are shown in Fig.~\ref{xtLIQUID}.
For even higher pressures the system solidifies, and the two
surfaces (A$_4$, b-tce-t-b$^{'}$-b) and (A$_3$, tce-ce-c$^{''}$-t-tce)
should continue
towards
a surface of
first-order
liquid-solid phase transitions (see A$_1$ in Fig.~\ref{3dschematic-fig}) which is not supported by the present model.
%
%
%

\section{Summary and Conclusion}

The phase diagram of the
general vectorized Blume-Emery-Griffiths
model has been explored within mean-field theory.
The model exhibits a liquid phase, which can be either a superfluid or a normal liquid,
and a vapor phase.
Depending on the choice of the coupling constants the model exhibits
various
topologies
of the phase diagram.
Here
we have focused on
those
topologies
of the phase diagram which are associated with
liquid-liquid phase transitions at coexistence with
the vapor phase.
Knowledge of them is a prerequisite for studying tricritical Casimir forces in $\Ht$ -$\Hf$ wetting films.
%
If the coupling constant
$j_s$,
which
facilitates
the occurrence of the superfluid phase, is turned off, the phase diagram is that of a normal binary liquid mixture (see Figs.~\ref{3dA}~and~\ref{t1}).
For nonzero but small values of
this
superfluid coupling constant the transitions to
the
superfluid phase are second order
only
(Figs.~\ref{3dB}~and~\ref{t2}). For larger values of this coupling constant,
the
transition to the
superfluid phase can also be of first order
(Figs.~\ref{3dC},~\ref{t3},~and~\ref{t4});
the liquid-liquid phase transitions
can be either between two normal liquids
or between superfluid and normal liquids. For even larger values of the superfluid coupling constant, the first-order liquid-liquid phase transitions occur only between
the superfluid and the normal fluid (Figs.~\ref{3dD}~and~\ref{t5}),
as it
is the case for
actual
$\Ht$ -$\Hf$ mixtures (see Figs.~\ref{he4}-\ref{3dschematic-fig}).

We conclude that
for a
suitable
set of coupling constants,
various
features of the
phase diagram of
$\Ht$ -$\Hf$ mixtures are captured
by the present approach (see Figs.~\ref{t5}-\ref{xtLIQUID}).
The detailed knowledge of the bulk phase diagram is necessary for studying wetting
phenomena
within the present model and, further, tricritical Casimir forces
acting on
wetting films.
The present model
lends itself also for investigations based on
Monte Carlo simulations.
This model of a binary liquid mixture incorporates vectorial degrees of freedom associated with the $\Hf$ particles
which covers the more complex behavior of the superfluid order parameter. It is interesting to note that
the sequence of the phase diagrams (Fig.~\ref{t2}(a), \ref{t3}(a), \ref{t4}(a), \ref{t5}(a)) exhibits the identical topologies as the phase diagrams of one-component dipolar fluids upon increasing
the dipole strength with the isotropic and ferromagnetic liquid corresponding to the normal liquid and the superfluid,
respectively~\cite{PhysRevLett.72.2422,PhysRevE.50.3814}.
For dipolar fluids the solid phase can be captured by off-lattice density functional theory~\cite{PhysRevE.54.1687}.

\end{document}